\documentclass[aps,prl,reprint,superscriptaddress,amsmath,amssymb,amsfonts,twocolumn]{revtex4-2}
\usepackage[utf8]{inputenc} 
\usepackage[english]{babel} 
\usepackage{microtype} 
\usepackage[dvipsnames]{xcolor} 
\usepackage{url} 
\usepackage{graphicx} 
\usepackage{bm} 
\usepackage{physics} 
\usepackage{mathtools} 
\usepackage{dsfont} 
\usepackage{mathrsfs} 
\usepackage{afterpage} 
\usepackage{times} 
\usepackage{algorithm}
\usepackage{algpseudocode}
\usepackage{tikz}
\usetikzlibrary{quantikz2}
\usepackage{hyperref} 
\hypersetup{
  colorlinks   = true, 
  urlcolor     = Blue, 
  linkcolor    = BrickRed, 
  citecolor   = PineGreen
}
\usepackage{cleveref} 

\usepackage{amsthm}         
\newtheorem{theorem}{Theorem}[section]
\newtheorem{lemma}[theorem]{Lemma}
\newtheorem{proposition}[theorem]{Proposition}

\date{\today}

\begin{document}
\title{Explicit quantum surrogates for quantum kernel models}

\author{Akimoto Nakayama}
\email{akimoto.nakayama@gmail.com}
\affiliation{%
  Graduate School of Engineering Science, Osaka University, 1-3 Machikaneyama, Toyonaka, Osaka 560-8531, Japan
}
\affiliation{
  Center for Quantum Information and Quantum Biology,
  Osaka University, 1-2 Machikaneyama, Toyonaka, Osaka 560-0043, Japan
}

\author{Hayata Morisaki}
\email{u748119d@ecs.osaka-u.ac.jp}
\affiliation{%
  Graduate School of Engineering Science, Osaka University, 1-3 Machikaneyama, Toyonaka, Osaka 560-8531, Japan
}

\author{Kosuke Mitarai}
\email{mitarai.kosuke.es@osaka-u.ac.jp}
\affiliation{%
  Graduate School of Engineering Science, Osaka University, 1-3 Machikaneyama, Toyonaka, Osaka 560-8531, Japan
}
\affiliation{
  Center for Quantum Information and Quantum Biology,
  Osaka University, 1-2 Machikaneyama, Toyonaka, Osaka 560-0043, Japan
}

\author{Hiroshi Ueda}
\email{ueda.hiroshi.qiqb@osaka-u.ac.jp}
\affiliation{
  Center for Quantum Information and Quantum Biology,
  Osaka University, 1-2 Machikaneyama, Toyonaka, Osaka 560-0043, Japan
}
\affiliation{Computational Materials Science Research Team, RIKEN Center for Computational Science (R-CCS), Kobe, Hyogo 650-0047, Japan
}

\author{Keisuke Fujii}
\email{fujii@qc.ee.es.osaka-u.ac.jp}
\affiliation{%
  Graduate School of Engineering Science, Osaka University, 1-3 Machikaneyama, Toyonaka, Osaka 560-8531, Japan
}
\affiliation{
  Center for Quantum Information and Quantum Biology,
  Osaka University, 1-2 Machikaneyama, Toyonaka, Osaka 560-0043, Japan
}
\affiliation{
  RIKEN Center for Quantum Computing (RQC),
  Hirosawa 2-1, Wako, Saitama 351-0198, Japan
}

\begin{abstract}
Quantum machine learning (QML) leverages quantum states for data encoding, with key approaches being explicit models that use parameterized quantum circuits and implicit models that use quantum kernels. 
Implicit models often have lower training errors but face issues such as overfitting and high prediction costs, while explicit models can struggle with complex training and barren plateaus.
We propose a quantum-classical hybrid algorithm to create an explicit quantum surrogate (EQS) for trained implicit models.
This involves diagonalizing an observable from the implicit model and constructing a corresponding quantum circuit using an extended automatic quantum circuit encoding algorithm. 
The EQS framework reduces prediction costs, provides a powerful strategy to mitigate barren plateau issues, and combines the strengths of both QML approaches.
\end{abstract}

\maketitle

\begin{figure*}[t]
 \centering
 \includegraphics[width=\linewidth]{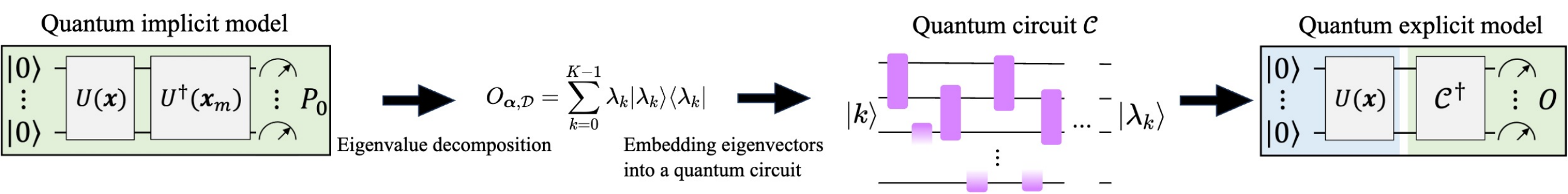}
 \caption{\textbf{
 Overview of the process to convert a trained implicit model to an explicit model (EQS).}
 An explicit model is constructed from a trained implicit model.
 First, we find the eigenvalues  $\lambda_k$ and eigenvectors $|\lambda_k\rangle$ of the observable $O_{\boldsymbol{\alpha}, \mathcal{D}}$ in Eq.~\eqref{eq:implicit_model_ob}. 
 Utilizing our extended AQCE algorithm, a quantum circuit $\mathcal{C}$ is constructed that satisfies the condition $\mathcal{C}|k\rangle\simeq|\lambda_{k}\rangle$ for $K$ eigenvectors $\{|\lambda_{k}\rangle\}_{k=0,...,K-1}$ with the accuracy desired by the user, where $|k\rangle$ is the computational basis.
 This yields an explicit model
 $\operatorname{Tr}\left[ \rho^\prime(\boldsymbol{x})  O \right]$, where $\rho^\prime(\boldsymbol{x})=\mathcal{C}^\dagger U(\boldsymbol{x})|\boldsymbol{0}\rangle\langle\boldsymbol{0}| U^{\dagger}(\boldsymbol{x}) \mathcal{C}$ is a density matrix and $O=\sum_{k=0}^{K-1} \lambda_k|k\rangle\langle k|$ is an observable.
 }
 \label{fig:overoview}
\end{figure*}

Quantum machine learning (QML) is an ambitious field that seeks to enhance machine learning capabilities by leveraging the power of quantum computers~\cite{Biamonte2017-ht,Cerezo2022-xy,dunjko2018machine,schuld2021machine}.
The ultimate goal is to demonstrate a ``quantum advantage'' by solving learning problems that are intractable for classical computers. 
A central and promising strategy toward this goal is the concept of a quantum feature map, which encodes data into high-dimensional quantum states~\cite{Havlicek2019-uo,Schuld2019-iz,Schuld2021-ma,Benedetti2019-ip,Cerezo2021-gk,Bharti2022-ie,Mitarai2018-rp}.
By representing data in this quantum feature space, QML models aim to uncover complex patterns and correlations that are beyond the reach of classical techniques.

Models using quantum features can be broadly classified into two categories: the explicit quantum models based on parameterized quantum circuits~\cite{Benedetti2019-ip,Cerezo2021-gk,Bharti2022-ie} such as quantum circuit learning~\cite{Mitarai2018-rp}, and the implicit models based on quantum kernels~\cite{Havlicek2019-uo,Schuld2019-iz,Schuld2021-ma}.
The former makes predictions about an input from the output of a single quantum circuit, which first embeds the input into a quantum state and then applies a parameterized quantum circuit to select important features stored in the state.
The model is therefore \textit{explicitly} specified via the description of the circuit used for prediction.
In contrast, the latter does so by calculating inner products of quantum features, that is, quantum kernel functions, among the data points and assessing them on a classical machine.
It is therefore \textit{implicit} in the sense that the quantum circuit itself does not describe the model. 
It has been demonstrated that both quantum models are capable of achieving good learning performance in benchmark tasks, as seen in numerical simulations~\cite{Schuld2019-iz,Schuld2020-jb,Liu2018-wm,Jerbi2021-oo,Skolik2022-sj,Nakayama2023-zc,D4DD00098F} and experiments using real quantum hardware~\cite{Havlicek2019-uo,Zhu2019-ss,Peters2021-so,Haug2023-bb,Nakayama2023-zc,Kusumoto2021-kd,Bartkiewicz2020-hx}.
Furthermore, in artificially and carefully designed scenarios, these models have demonstrated rigorous advantages over traditional classical methodologies~\cite{Liu2021-pe,Huang2021-to,Pirnay2022-si,Jerbi2021-oo,Dunjko2017-ra}.

The choice between these two paradigms involves a nuanced trade-off. 
While implicit models can find an optimal solution to minimize empirical loss due to the Representer Theorem~\cite{Smola1998-zt,Scholkopf2002-mh}, both approaches face fundamental challenges rooted in the curse of dimensionality~\cite{cerezo2023does}.
For fidelity kernels, this can lead to poor generalization from exponential concentration~\cite{thanasilp2024exponential}. 
Conversely, explicit models face significant training challenges due to non-convex landscapes and the barren plateau (BP) phenomenon~\cite{McClean2018-sg,Cerezo2021-gg,Larocca_2025}, as well as the difficulty of identifying a suitable ansatz.
Although both concentration and BPs stem from the same root cause, their impact on trainability is distinct, as we detail in Appendix~\ref{sec: relation}. 
Indeed, a well-trained implicit model is not guaranteed to outperform an explicit one on all tasks~\cite{Jerbi2023-lq}.

Acknowledging this complex landscape, our work addresses a specific and practical scenario. 
Since the ``curse of dimensionality'' renders any learning algorithm intractable on unstructured data, our work—like most successful machine learning—focuses on ``learnable'' problems where the data possesses sufficient structure for a global fidelity quantum kernel to be trained successfully. 
This success, however, reveals a critical bottleneck, i.e., the $O(M)$ prediction cost.

The central question our work addresses is therefore how we can systematically convert such a ``successful-but-slow'' implicit model into a ``fast-and-practical'' explicit model, preserving its high performance while achieving an $O(1)$ prediction cost.

In this work, we propose a quantum-classical hybrid algorithm to construct a quantum explicit model that acts as a {\it surrogate} for a trained quantum implicit model, which we call the explicit quantum surrogate (EQS), as shown in Fig.~\ref{fig:overoview}.
Our algorithm is composed of two key steps. 
The first step is the scalable diagonalization of an observable determined by the trained implicit model.
An important observation is that this observable can be diagonalized with $\mathrm{poly}(M)$ resources, despite the $2^n$ dimension of the feature space, and we find numerically that it is often low-rank approximable.
The second step is the construction of a quantum circuit that approximately diagonalizes the observable.
For this purpose, we extend the automatic quantum circuit encoding (AQCE) algorithm~\cite{Shirakawa2021-uj}.
Our extension enables the generation of a circuit $\mathcal{C}$ that creates an isometry for diagonalizing the low-rank observable, and importantly, does so without assuming any fixed ansatz structure.
This constructed circuit allows us to evaluate the expectation value by simply measuring the resulting quantum states in the computational basis.

This two-step construction process, while powerful, constitutes the main source of the one-time resource cost for our framework.
A key contribution of this work is therefore a comprehensive analysis of these required resources, including measurement shots, classical processing, and circuit depth scalability.
As detailed in Appendix~\ref{sec:cost_analysis}, we show this cost is a justifiable investment for applications that require high prediction throughput, thanks to the $O(1)$ prediction efficiency of the resulting EQS.

This procedure allows us to construct an explicit model that has a training loss almost equivalent to the trained implicit model.
Furthermore, the generated EQS provides insights about ansatz structures suitable for specific datasets and can be used as a high-quality starting point to mitigate the BP phenomenon, as it does not depend on a randomly initialized ansatz.

We evaluate the performance of EQS for classification tasks through numerical simulations. 
The prediction accuracy of EQS using the MNISQ dataset~\cite{Placidi2023-kq} (a 10-qubit quantum circuit dataset with $M=$10,000 data points) and the VQE-generated dataset~\cite{Nakayama2023-zc} (a 12-qubit quantum circuit dataset with $M=$1,800 data points) decreases by less than 0.010 compared to trained implicit models.
It is notable that these results are obtained by only considering $K\leq 10$ eigenvectors.
Additionally, we find that, even when the isometry generated by AQCE has a fidelity of only $0.6$, the decrease in prediction accuracy remains at the same level.
This suggests that we do not need to optimize the circuit carefully, and the computational cost for optimizing the circuit construction step may be smaller than one might expect.
Furthermore, we demonstrate that our method effectively mitigates the BP phenomenon. 
A scaling analysis reveals that the EQS initialization suppresses the exponential decay of gradients observed in randomly initialized circuits, with the performance gap widening to more than 5 orders of magnitude at 16 qubits.
From these observations, we believe that our proposal is a valuable tool not only for reducing the prediction cost of implicit models but also for potentially enhancing the trainability of explicit models.
Our framework is positioned in contrast to alternative strategies, such as those that modify the kernel itself or create classical surrogates, as detailed in Appendix~\ref{sec: compare_other_method}.

\emph{Preliminary.}---
Let us first define the notations and terms, which are mostly based on those used in Refs.~\cite{Jerbi2023-lq, Schuld2021-ma}. 
Let $\rho(\boldsymbol{x})=|\psi(\boldsymbol{x})\rangle\langle\psi(\boldsymbol{x})|=U(\boldsymbol{x})|0\rangle\langle0| U^{\dagger}(\boldsymbol{x})$ be an $n$-qubit quantum feature state that encodes an input $\boldsymbol{x}$ generated via a feature-encoding quantum circuit $U(\boldsymbol{x})$. 
We define the explicit model as:
\begin{align}
f_{\mathrm{explicit}}(\bm{x};\bm{\theta}) = \mathrm{Tr}[\rho(\bm{x})V(\boldsymbol{\theta})^{\dagger} O V(\boldsymbol{\theta})],
\end{align}
where $O$ is an efficiently measurable operator and  $V(\boldsymbol{\theta})$ is a parameterized quantum circuit with trainable parameters $\boldsymbol{\theta}$.
The training of explicit models is performed by optimizing the parameters $\boldsymbol{\theta}$ to minimize empirical loss.
The implicit models are defined as:
\begin{align}
    f_{\mathrm{implicit}}(\bm{x};\bm{\alpha}) = \sum_{m=1}^M\alpha_m\mathrm{Tr}[\rho(\bm{x}_m)\rho(\bm{x})]
\end{align}
where $\{\boldsymbol{x}_{m}\}=:\mathcal{D}$ is the training dataset, and $\boldsymbol{\alpha}\in\mathbb{R}^M$ is the model parameter which is determined through training.
An important observation that we make use of in this work is that $f_{\mathrm{implicit}}$ can be rewritten in the form of $f_{\mathrm{implicit}}(\bm{x};\bm{\alpha}) = \mathrm{Tr}[O_{\boldsymbol{\alpha}, \mathcal{D}}\rho(\bm{x})]$ by defining
\begin{align}
O_{\boldsymbol{\alpha}, \mathcal{D}}=\sum_{m=1}^{M} \alpha_{m} \rho\left(\boldsymbol{x}_{m}\right). \label{eq:implicit_model_ob}
\end{align}

\emph{Algorithm to construct explicit quantum surrogate.}---
Our algorithm consists of two main components.
The first component is the eigenvalue decomposition of the observable $O_{\boldsymbol{\alpha}, \mathcal{D}}$ of the trained implicit model.
The second component is constructing a quantum circuit that produces these eigenvectors.
Although there are many ways of achieving this, our choice is to employ the AQCE~\cite{Shirakawa2021-uj} extended for constructing an isometry that allows us to generate a circuit without assuming a fixed ansatz.
The overview of our proposed algorithm is shown in Fig.~\ref{fig:overoview}.
We describe each component in the sequence below.

First, we perform training of the implicit model and diagonalize the observable $O_{\boldsymbol{\alpha}, \mathcal{D}}$.
As $\mathcal{S}=\operatorname{span}\{|\psi(\boldsymbol{x}_m)\rangle\}$ is an invariant subspace of 
$O_{\boldsymbol{\alpha}, \mathcal{D}}$, 
it is sufficient to diagonalize $O_{\boldsymbol{\alpha}, \mathcal{D}}$ within this subspace.
For example, we can employ the following process to achieve this.
We first determine the set of orthogonal basis vectors $\{|e_i\rangle\}_{i=1}^{\operatorname{dim} (\mathcal{S})}$ of $\mathcal{S}$ using e.g. the Gram-Schmidt process.
The inner products $ \langle\psi(\bm{x}_m)|\psi(\bm{x}_{m^\prime})\rangle$ required in the process can be obtained using the Hadamard test~\cite{Schuld2018-eg}.
Next, we compute the matrix elements of $O_{\boldsymbol{\alpha}, \mathcal{D}}$ with respect to the new basis $\{|e_i\rangle\}$.
They can be calculated as 
\begin{align}
    [O_{\boldsymbol{\alpha}, \mathcal{D}}]_{ij}:= \langle e_i|O_{\boldsymbol{\alpha}, \mathcal{D}}|e_j \rangle 
    = \sum_{m=1}^{M} \alpha_m \langle e_i|\rho(\bm{x}_m)|e_j \rangle.
\end{align}
We can then diagonalize it classically to obtain its eigenvalues $\{\lambda_k\}_{k=0}^{\operatorname{dim} (\mathcal{S})-1}$ and corresponding eigenvectors $\{|\lambda_{k}\rangle\}_{k=0}^{\operatorname{dim} (\mathcal{S})-1}$ expressed as linear combinations of $\ket{\psi(\bm{x}_m)}$. 
Using these, the implicit model can now be rewritten as $f_{\mathrm{implicit}}(\boldsymbol{x};\bm{\alpha}) = \sum_{k=0}^{\operatorname{dim} (\mathcal{S})-1} \lambda_k \bra{\lambda_k}\rho(\bm{x})\ket{\lambda_k}$.
Note that it is often possible to truncate the sum at $K\ll M$ without significant performance decrease, as we will discuss later and show in the numerical experiments.
The validity of this truncation is not merely a heuristic; as we demonstrate with a formal error bound and an empirical spectral analysis in Appendix~\ref{sec:low_rank_justification}, it is a well-grounded strategy for typical, learnable datasets.
Truncation retains important information in a model while reducing its complexity. 
It may offer advantages in mitigating overfitting, but we leave such an analysis as future work.

Next, we construct a quantum circuit $\mathcal{C}$ that satisfies $\mathcal{C}|k\rangle\approx |\lambda_{k}\rangle$ for $k=0,...,K-1$, where $|k\rangle$ denotes the computational basis states.
For this purpose, we extend AQCE to isometries.
The original AQCE is an algorithm that generates a circuit that outputs a target state $|\Psi\rangle$ with the desired accuracy. 
It constructs quantum circuits by iteratively adding two-qubit unitary gates without assuming any fixed ansatz.
The optimization is performed in a manner similar to that in tensor network methods~\cite{Evenbly2009-cq}.
A brief review of AQCE and the extensions done in this work are given in Appendix~\ref{sec: aqce} and \ref{sec: extended_aqce}, respectively.

A key strength of our EQS framework is its robustness to the potential sub-optimality of the circuit construction step. 
While the AQCE algorithm is powerful, it must navigate a highly complex, non-convex optimization landscape, and thus provides no guarantee of finding a global optimum. 
Our numerical results demonstrate, however, that a perfect circuit is not required to build a high-performance EQS.

The implicit model can now be translated into an explicit model, that is, an EQS: 
\begin{align}\label{eq:eqs}
f_{\mathrm{EQS}}(\boldsymbol{x}) = \mathrm{Tr}[O\mathcal{C}^\dagger\rho(\bm{x})\mathcal{C}] \simeq f_{\mathrm{implicit}}(\bm{x};\bm{\alpha}),
\end{align}
where $O=\sum_{k=0}^{K-1}\lambda_{k}|k\rangle \langle k|$.
This explicit formulation offers a significant advantage in prediction efficiency, not only in terms of circuit executions but also in statistical cost. 
While a prediction requires only a single type of circuit execution, a potential concern is the number of measurement shots needed for estimation. 
However, as we prove in Appendix~\ref{sec: prediction_cost}, the required sample complexity is determined by the properties of the observable $O$ and is independent of the system size $n$. 
This addresses a key challenge for the practical application of QML models and rigorously establishes the efficiency of the EQS prediction phase.

It should be noted that the above process of generating EQS can be understood as quantum architecture search (QAS) for finding a well-performing circuit for explicit models~\cite{noauthor_undated-wm}.
We present an example of the quantum circuit structure found in our numerical simulations, described later in Appendix~\ref{fig:heat1}.
The analysis of the quantum circuit structure found by EQS is an interesting direction to explore, but it is beyond the scope of this work and will be considered in future research.

The possibility of low-rank approximation significantly affects cost of the AQCE step.
Even though it does not raise intrinsic exponential cost to the number of qubits within the iterations for circuit optimization, we would expect that the AQCE would become increasingly difficult when the number of vectors to be constructed, $K$, is large.
The effectiveness of a low-rank approximation can be guaranteed when $\operatorname{dim}(\mathcal{S})$ is small, which we argue here to be expected for quantum features $\ket{\psi(\bm{x})}$ that are well-designed, in the sense that ``similar'' $\bm{x}$ are mapped to similar feature vectors $\ket{\psi(\bm{x})}$.
For supervised learning with input $\bm{x}$ and output $y$, we say data $\bm{x}$ and $\bm{x}'$ are similar when corresponding $y$'s are equal or close.
For such well-designed features, we can expect that a large portion of $\{\ket{\psi(\bm{x}_m)}\}$ is linearly dependent, thus making $\operatorname{dim}(\mathcal{S})$ small.
Such well-designedness also guarantees prediction performance. Ref.~\cite{Huang2021-to} (Eq.~(8)) shows that, if we wish to predict $y$ in the form of $y=\mathrm{Tr}[A\rho(\bm{x})]$ for an unknown observable $A$ having a sufficiently small norm, $\mathrm{Tr}(A^2)$, we can achieve a prediction error of $\mathcal{O}(\operatorname{dim}(\mathcal{S}))$.
We therefore assume that for a well-designed feature that predicts well, the observable of the trained implicit model, $O_{\boldsymbol{\alpha}, \mathcal{D}}$, admits a low-rank approximation.

\emph{Numerical analysis of EQS performance.}---
To evaluate our algorithm, we derive EQS from the implicit model of the support vector machine (SVM)~\cite{Pl1999,Wu2003-nv} using global fidelity quantum kernel~\cite{Havlicek2019-uo,Schuld2019-iz}.
We then compare the classification accuracies using two datasets: the MNISQ dataset~\cite{Placidi2023-kq}, which contains 10,000 samples with 10 labels, and the VQE-generated dataset~\cite{Nakayama2023-zc}, which contains 1,800 samples with 6 labels.
The MNISQ dataset is derived from the quantum encoding of the MNIST dataset~\cite{Lecun1998-dr} and consists of labeled quantum circuits with 10 qubits.
The VQE-generated dataset, created using the VQE algorithm~\cite{Peruzzo2014-gz,Tilly2022-nh}, comprises labeled quantum circuits with 12 qubits and has the property that output states of circuits with different labels exhibit extremely low fidelity.
A detailed description of these datasets, including their origin and construction, is provided in Appendix~\ref{sec: dataset_descriptions}.
They are provided in QASM format~\cite{Cross2022-pb}.
Circuits in these datasets are essentially labeled according to the similarity of the output states; those with the same label output similar quantum states.
The QASM string is an input data $\bm{x}$ in this case, and we simply choose $U(\bm{x})$ as the circuit described by the string $\bm{x}$.

To assess how low-rank approximations affect classification accuracy, we compare multiple EQSs, each with a different value of $K$.
In addition, to evaluate how approximations arising within the use of our extended AQCE algorithm for generating isometries affect classification accuracy, we compare two different models.
The first model is an EQS represented by Eq.~\eqref{eq:eqs} optimized by our algorithm until it satisfies $F^{(k)}:=\left|\left\langle k\left|\mathcal{C}^{\dagger}\right| \lambda_k\right\rangle\right|>0.6$ for all $k$.
The second model, which we call the exact low-rank model, directly uses the results of low-rank approximations of $O_{\boldsymbol{\alpha}, \mathcal{D}}$, thus avoiding the error induced by the AQCE step.
We adopt the one-vs-rest strategy for multi-class classification; that is, we construct a multi-class classifier by combining multiple binary classifiers.
Note that this requires us to construct $O_{\bm{\alpha}, \mathcal{D}}$ and the corresponding circuit $\mathcal{C}$ defining EQS for each label $l$, which we denote by $O_{\bm{\alpha}, \mathcal{D}}^{(l)}$ and $\mathcal{C}^{(l)}$.
The accuracy shown in the results is defined as the number of correct classifications divided by the total number of test data points.
The quantum circuit simulation is performed using Qulacs 0.5.6~\cite{Suzuki2021-nj} in a noiseless environment.
While the main results are presented for a noiseless environment to clearly assess the core performance, a detailed analysis of the method's robustness against realistic shot noise is provided in Appendix~\ref{sec:cost_analysis}. 
This analysis shows that the impact on both final accuracy and, crucially, on the construction cost is limited and manageable.
We detail the experimental conditions in Appendix~\ref{sec: condition} and show the results in Fig.~\ref{fig:sim_result}.
The results for other datasets can be found in Appendix~\ref{sec: add_simu}, which are consistent with the results presented here.

\begin{figure}[t]
 \centering
\includegraphics[width=\linewidth]{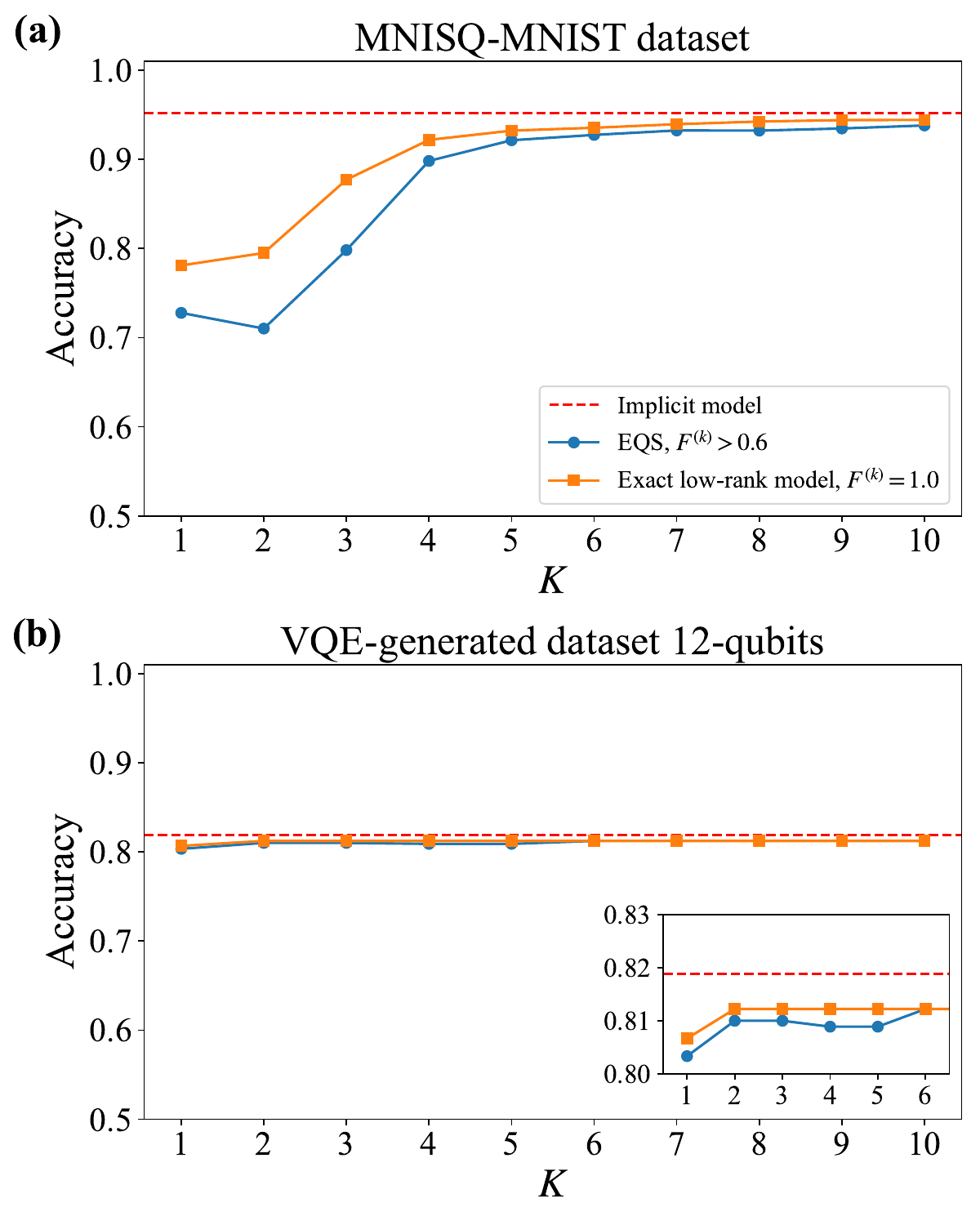}
 \caption{
 \textbf{Performance of EQS on MNISQ-MNIST and 12-qubit VQE-generated dataset.}
 The vertical axis represents the classification accuracy on the test data.
 The horizontal axis represents the number of eigenvectors $K$ used in the eigenvalue decomposition of $O_{\boldsymbol{\alpha},\mathcal{D}}$.
 The EQS refers to Eq.~\eqref{eq:eqs} with fidelities $F^{(k)}>0.6$ for all $k$.
 The exact low-rank model is obtained by exact low-rank approximations of $O_{\boldsymbol{\alpha}, \mathcal{D}}$, which is equivalent to Eq. \eqref{eq:eqs} with $F^{(k)}=1.0$ for all $k$.
 An inset in Fig.~\ref{fig:sim_result} (b) provides a detailed, magnified view of a specific area depicted in this panel.
 }
 \label{fig:sim_result}
\end{figure}

First, we discuss the behavior of the exact low-rank models.
In Fig.~\ref{fig:sim_result} (a) and (b), we observe that their accuracy improves and approaches the accuracy of the original one as $K$ increases.
For instance, with the MNISQ-MNIST dataset as shown in Fig.~\ref{fig:sim_result} (a), the exact low-rank model exhibits only a 0.010 decrease in accuracy compared to the implicit model at $K=10$.
For the VQE-generated dataset in Fig.~\ref{fig:sim_result} (b), there is a mere 0.014 decrease in accuracy compared to the implicit model at $K=1$.
This can be explained by the fact that $\ket{\psi(\bm{x})}$ with different labels exhibit extremely low fidelity in this dataset.
Therefore, looking at the fidelity between a state $\ket{\psi(\bm{x})}$ for an unknown $\bm{x}$ and an average of $\ket{\psi(\bm{x}_m)}$ within the same label $l$ in a training set would be sufficient to classify $\bm{x}$.
Indeed, the first eigenvector of $O_{\bm{\alpha}, \mathcal{D}}^{(l)}$ has a high fidelity of over 0.7 with $\ket{\psi(\bm{x})}$ belonging to the same label $l$ for most $l$'s, as shown in Fig.~\ref{fig:add_analysis_a} of Appendix~\ref{sec:additional_analysis}.

Overall, results indicate that high accuracy can be achieved with $K\ll M$ and that low-rank approximation is effective, as $M$ is on the order of $10^3-10^4$ for each dataset.
The effectiveness of this approximation can also be seen from the mean value of the cumulative contribution ratio $\frac{\sum_{i=0}^{K-1} \lambda_i^2}{\sum_{i=0}^{M-1} \lambda_i^2}$, which is shown in Fig.~\ref{fig:add_analysis_b} of Appendix~\ref{sec:low_rank_justification}.
For the MNISQ-MNIST and VQE-generated datasets, the values are 0.798 at $K=10$ and 0.744 at $K=6$, respectively.
We believe that this high amenability to low-rank approximation is reasonable because observables are constructed from linear combinations of the quantum features that encode training data, and the quantum features in this example are well-designed in the sense that data with the same label are mapped to similar quantum states.

Next, we discuss the performance of the EQS models.
Fig.~\ref{fig:sim_result} (a) and (b) show that the impact of approximations made in the AQCE step on accuracy is surprisingly small.
The decrease in accuracy from the exact low-rank model (which assumes a perfect isometry with fidelities of 1.0) is only $0.008$ for the MNIST dataset at $K=10$.
No decrease is observed for the VQE-generated dataset at $K=6$.
It should be emphasized that the condition imposed on fidelities is only $F^{(k)}>0.6$.
This finding has significant practical implications for the resource cost of EQS construction.
It reveals a favorable trade-off between circuit depth and prediction accuracy: by accepting a minor compromise in fidelity, which is sufficient for high accuracy, we can halt the resource-intensive AQCE algorithm much earlier.
This directly translates to a significantly shallower circuit, mitigating one of the key construction costs detailed in Appendix~\ref{sec:cost_analysis}.
We assume this is because the use of imperfect replication of the eigenvectors could change the decision boundaries established in the training phase, but did not affect the prediction results due to the margin preserved by the SVM.
Given that the currently available quantum computers are affected by a non-negligible amount of noise, the fact that low target fidelity results in minimal degradation of accuracy may offer an advantage when executing our method on actual devices.

\begin{figure}[t]
  \centering
  \includegraphics[width=\linewidth]{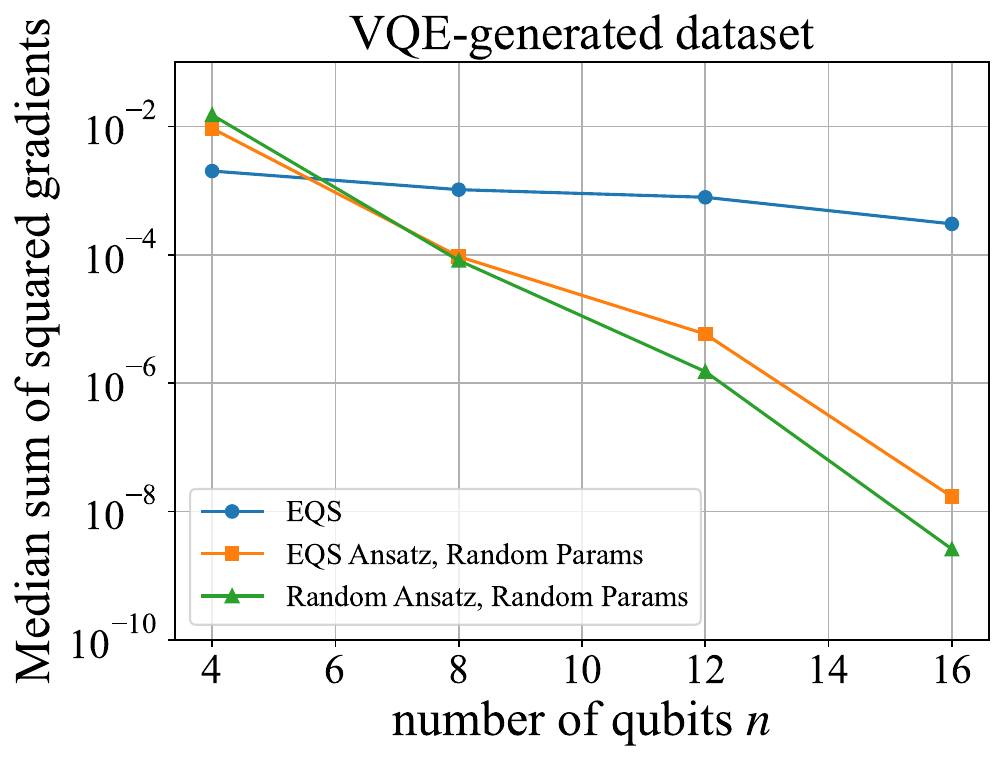}
  \caption{
    \textbf{Median sum of squared gradients for explicit models with different initializations.}
    For each target label, we compute the sum of squared gradients at the first training step. 
    The horizontal axis indicates the number of qubits $n$.
    The vertical axis shows the median of these values across the target labels. 
    For $n<16$, all six labels were used; for the $n=16$ point, a subset of four labels was used due to the high simulation cost.}
  \label{fig:grad_result}
\end{figure}

To further validate the practical advantage of our approach, we also benchmarked EQS against simpler heuristic classifiers. 
These results, detailed in Appendix~\ref{sec: baseline_comparison}, confirm that for complex tasks where the original kernel model excels, our EQS significantly outperforms these baselines, justifying the necessity of its more sophisticated construction.

\emph{EQS as initialization strategy.}---
Our strategy to construct the EQS via AQCE automatically finds a circuit $\mathcal{C}$ that defines a well-performing explicit model.
It is, therefore, natural to use the circuit found by our algorithm as an initial point for training an explicit model.
However, it is not clear if this strategy provides a trainable initialization, that is, non-vanishing gradients.
Here, we compare the gradients of the cross-entropy loss function when the explicit models are randomly initialized and when they are initialized to an EQS.
To this end, we first construct the EQS, which is identical to that employed in the preceding numerical analysis, and then compute the loss gradients on an independent test set.
The gradients are computed by regarding the two-qubit unitary gates in $\mathcal{C}^{(l)}$ as parameterized arbitrary two-qubit unitary gates with $15$ parameters.
To precisely identify the advantage of our strategy, we compare the gradients under three distinct initialization schemes for each label. 
The first is our EQS-initialized model, using both the circuit structure and parameters found by our method. 
The second uses the same EQS-found ansatz, but with its parameters randomly initialized in $[0,2\pi)$.
This serves as a crucial control to test if the ansatz structure alone is beneficial. 
The third is a baseline using a randomly structured ansatz of similar depth, also with random parameters.
The quantum circuit simulation is performed using Qulacs 0.5.6~\cite{Suzuki2021-nj} under a noiseless environment.
The details of the experimental conditions are described in Appendix~\ref{sec: condition}.

We present the results of our scaling analysis in Fig.~\ref{fig:grad_result}, which shows the sum of squared gradients for each of the three initialization schemes as a function of system size. 
The results provide clear evidence for BP mitigation. 
We observe that the model using the EQS-found ansatz with random parameters and the model with a random ansatz both suffer from the characteristic exponential decay of gradients. In contrast, the EQS-initialized model completely suppresses this trend.
This highlights that the circuit structure alone is insufficient to avoid BPs; the choice of EQS-found initial parameters is the critical component. 
The practical impact of this mitigation is stark: the performance gap between the EQS initialization and its randomly initialized counterpart widens exponentially with system size, reaching a difference of more than 5 orders of magnitude at 16 qubits.
A significant gradient enhancement was also observed for the 10-qubit MNISQ-MNIST dataset, suggesting this effect is not specific to the VQE-generated dataset (see Appendix~\ref{sec: add_simu}).
These findings establish EQS as a powerful strategy for mitigating trainability issues in explicit models.

\emph{Conclusion.}---
In this work, we have introduced and validated the explicit quantum surrogate (EQS) framework, a systematic method to convert a trained, high-performance implicit model into a fast and practical explicit model.
We have shown that EQS achieves a prediction cost of $O(1)$, a significant improvement over the $O(M)$ cost of kernel methods, while maintaining a classification accuracy comparable to the original model.
Furthermore, we demonstrated that using EQS as an initialization strategy provides a high-quality starting point for explicit model training, yielding initial gradients that are orders of magnitude larger than a random initialization, thereby offering a path to mitigate the barren plateau problem.

We position this work as a proof-of-concept for a new hybrid QML paradigm.
We acknowledge that the one-time construction cost of EQS, particularly the quantum resources required for the circuit-building step, presents a challenge for current near-term (NISQ) devices.
However, we believe this initial investment can be justified in application domains where high prediction throughput is required.
Moreover, as fault-tolerant quantum computers become available and more efficient circuit construction algorithms are developed, we anticipate that this overhead will become less of a bottleneck, further broadening the applicability of the EQS approach.

The EQS framework opens several intriguing avenues for future research.
One key direction is to explore further training of the EQS, which could potentially improve generalization and mitigate overfitting issues of the original kernel model.
Another is to investigate the physical or problem-specific meaning of the circuit structures discovered by EQS.
A third avenue involves extending the framework to other kernel types, such as projected quantum kernels~\cite{Huang2021-to,gan2023unified}, which may allow tackling a broader class of problems by leveraging specific inductive biases (see Appendix~\ref{sec: applicability_projected_kernels} for a detailed discussion).
Finally, the high inference efficiency of EQS models makes them promising candidates for deployment in novel computational settings, such as real-time response systems, inference on large static datasets, or QML on resource-constrained edge devices, stimulating new connections between QML and practical computer science applications.

K.M. is supported by JST PRESTO Grant No. JPMJPR2019 and JSPS KAKENHI Grant No. 	23H03819.
H.U. is supported by JSPS KAKENHI Grant No. JP21H04446, JP21H05182, JP21H05191, JST CREST Grant No. JPMJCR24I1, and the COE research grant in computational science from Hyogo Prefecture and Kobe City through Foundation for Computational Science.
This work is supported by MEXT Quantum Leap Flagship Program (MEXT Q-LEAP) Grant No. JPMXS0120319794, JST COI-NEXT Grant No. JPMJPF2014, and JST CREST JPMJCR24I3.

\bibliography{2_main}

\clearpage
\appendix
\onecolumngrid
\begin{center}
\large{\textbf{Supplemental material for ``Explicit quantum surrogates for quantum kernel models''}}
\end{center}
\renewcommand{\thefigure}{S\arabic{figure}} 
\renewcommand{\thealgorithm}{S\arabic{algorithm}} 
\setcounter{figure}{0}
\setcounter{equation}{0}

\section{Comparison with alternative methods for prediction acceleration}
\label{sec: compare_other_method}
In this section, we provide a detailed comparison between our explicit quantum surrogate (EQS) framework and other notable methods that aim to reduce the prediction cost associated with quantum kernel models. 
We focus on two main classes of alternatives: other quantum kernel approaches with improved efficiency, such as linear projected quantum kernels (LPQKs)~\cite{Huang2021-to,Kubler2021-du,gan2023unified}, and methods that construct classical surrogate models~\cite{Schreiber2023-di,Jerbi2023-wt}.

\subsection{vs. Linear projected quantum kernels (LPQKs)}
LPQKs~\cite{Huang2021-to,Kubler2021-du,gan2023unified} are a family of quantum kernels designed to be efficient by restricting measurements to local subsystems. 
This introduces an inductive bias, which contrasts with the global fidelity quantum kernel used by EQS.
The choice between EQS and LPQKs involves a series of trade-offs.

\begin{itemize}
    \item \textbf{Training and prediction costs:} 

    A key trade-off hinges on the problem's locality, denoted by subsystem size $S$. 
    The sampling cost for training the underlying global fidelity quantum kernel, which EQS is designed to surrogate, scales as  $O\left(M^2\right)$, versus $O\left(M \cdot 3^S\right)$ for LPQKs, where $M$ is the number of training data points~\cite{gan2023unified}.
    The prediction cost for EQS is $O(1)$, while for LPQKs it is $O(3^S)$. 
    Therefore, for problems with a local structure corresponding to a small subsystem size $S$, LPQKs is more efficient.
    However, for non-local problems where a large $S$ is required, the exponential scaling makes LPQK's cost prohibitive, and the polynomial scaling of the global fidelity kernel approach becomes advantageous. 
    
    In terms of circuit depth, LPQKs are also efficient. 
    They require only a single implementation of the encoding circuit. 
    Estimating the global fidelity quantum kernel, in contrast, requires either doubling the circuit depth via the inversion test~\cite{hubregtsen2022training} or doubling the qubit count via the SWAP test~\cite{hubregtsen2022training}.

    \item \textbf{Expressivity vs. Inductive bias:}

    EQS begins with the global fidelity quantum kernel, which is maximally expressive as it is equally sensitive to all $4^n$ orthogonal basis observables (e.g., the set of all $n$-qubit Pauli strings), imposing no a priori structural bias on the learning problem~\cite{gan2023unified}.
    LPQKs, by contrast, are intentionally less expressive, imposing an inductive bias that prioritizes local information. 
    This represents a fundamental conceptual difference: EQS pursues a ``high-power-then-compress'' strategy, while LPQKs follows a ``restricted-from-the-start'' approach.

    \item \textbf{Adaptability and fine-tuning:}

    A key point of divergence lies in their adaptability.
    An LPQKs model, even when made explicit by solving the primal problem, consists of a fixed feature map with tunable linear weights.
    Fine-tuning is thus restricted to finding a new decision boundary within a static feature space. 
    EQS provides two levels of adaptability: a lightweight update of the classical weights $\lambda_k$ in its observable $O=\sum_{k=0}^{K-1} \lambda_k|k\rangle\langle k|$, and a more flexible update of the gate parameters $\boldsymbol{\theta}$ within the EQS circuit $\mathcal{C}$ that reshapes the feature space itself.
    This makes EQS potentially more robust to significant concept drifts.
\end{itemize}

\subsection{vs. Classical surrogate models}
Another approach to reducing cost is to construct a purely classical surrogate for the quantum model. 
Some methods aim to dequantize the full workflow, from training to prediction~\cite{Landman2022-ui,Sweke_2025, sahebi2025dequantization}. 
Our work, however, is motivated by scenarios where quantum training is necessary--for instance, when tackling problems where classical models face fundamental limitations in achieving a quantum advantage. 
As such, these full dequantization strategies fall outside our primary scope.

We instead focus our comparison on the more directly relevant task of creating classical surrogates for the prediction phase only, after a model has been trained on a quantum computer.
However, current techniques for this have their own limitations that define the niche where EQS is particularly advantageous. 
Techniques based on random fourier features (RFF)~\cite{Schreiber2023-di} are typically restricted to shift-invariant kernels, and their sample complexity is known to scale exponentially with input data dimension.

Similarly, shadowfied flipped models~\cite{Jerbi2023-wt} have a construction cost that scales with the locality of the target observables.
This presents a challenge for models like ours that start with a global fidelity quantum kernel, which can learn highly non-local or complex observables. 
For such models, the cost of building an accurate classical surrogate can become intractable.

Furthermore, it has been proven that there exist quantum models that cannot be efficiently ``dequantized'' for prediction~\cite{Jerbi2023-wt}. 
EQS is intentionally designed as a quantum surrogate to fill this gap. 
It provides an efficient prediction pathway for these challenging, ``non-dequantizable'' scenarios where a quantum solution is required, while also retaining the unique, two-level adaptability of a quantum circuit model.

\section{On barren plateaus and the EQS mitigation strategy}
This appendix provides a brief review of the barren plateau phenomenon and contextualizes our proposed EQS strategy. 
The following discussion on mitigation strategies is largely based on the comprehensive review presented in~\cite{Larocca_2025}, and the literature cited therein.

\subsection{The barren plateau phenomenon}
A central challenge hindering the trainability of variational quantum algorithms is the Barren Plateau (BP) phenomenon~\cite{Larocca_2025}. 
This refers to the concentration of the loss function's gradient, where its variance vanishes exponentially with the number of qubits $n$.
Intuitively, this means that for a sufficiently large system size, the optimization landscape becomes almost entirely flat and featureless. 
As a result, gradient-based optimizers cannot find a useful descent direction, leading to training stagnation. 
This requires an exponential number of measurement shots to determine the direction needed to minimize the cost function.

More formally, let's consider a loss function $\ell_{\boldsymbol{\theta}}(\rho, O)$ defined by an initial state $\rho$ and an observable $O$, which depends on randomly chosen circuit parameters $\boldsymbol{\theta}$.
The BP phenomenon means that for large systems, both the loss function's value and its gradients concentrate around their average values.
Specifically, the probability that the loss deviates from its expectation value by more than a small constant $\delta>0$ (where $\delta \in \Omega(1 / \operatorname{poly}(n))$ ) is exponentially suppressed with the number of qubits $n$ :

\begin{equation}
\operatorname{Pr}_{\boldsymbol{\theta}}\left(\left|\ell_{\boldsymbol{\theta}}(\rho, O) - \mathbb{E}_{\boldsymbol{\theta}}\left[\ell_{\boldsymbol{\theta}}(\rho, O)\right]\right| \ge \delta\right) \in \mathcal{O}\left(\frac{1}{b^n}\right),
\end{equation}

where $b>1$ is a constant. 
This exponential concentration also holds for any partial derivative $\partial_\mu \ell_{\boldsymbol{\theta}}(\rho, O)$:

\begin{equation}
\operatorname{Pr}_{\boldsymbol{\theta}}\left(\left|\partial_\mu \ell_{\boldsymbol{\theta}}(\rho, O) - \mathbb{E}_{\boldsymbol{\theta}}\left[\partial_\mu \ell_{\boldsymbol{\theta}}(\rho, O)\right]\right| \ge \delta\right) \in \mathcal{O}\left(\frac{1}{b^n}\right).
\end{equation}

The fundamental cause of this phenomenon is the ``curse of dimensionality''~\cite{cerezo2023does}; the Hilbert space that the parameterized quantum circuit must explore is exponentially large in the number of qubits. 
This leads to the expressive, yet unstructured, ansatz effectively behaving like random unitary operations, causing the concentration of measure effects that result in BPs.

However, it is crucial to note that the absence of BPs is a necessary, but not sufficient, condition for gradient-based trainability~\cite{gil2024relation}.
The condition is necessary because gradient-based optimizers fundamentally rely on the existence of non-vanishing gradients almost everywhere in the landscape to find a path toward a solution~\cite{heaton2018ian}.
At the same time, it is not a sufficient condition because the absence of plateaus does not preclude other challenging landscape features~\cite{you2021exponentially,anschuetz2022quantum}.

Therefore, while mitigating BPs is a critical first step, the global structure of the optimization landscape must also be considered for developing truly effective variational quantum algorithms.

\subsection{Mitigation via informed initialization strategies}
Theoretical analyses of BPs are often predicated on the assumption of random parameter initialization.
While this assumption is useful for understanding the average-case behavior of an ensemble of circuits, it contrasts with practical application, where it is widely recognized that a random starting point is rarely a viable strategy. 
Consequently, a key mitigation strategy is to abandon naive random approaches in favor of ``informed'' initialization methods~\cite{Larocca_2025}. 
The goal of these strategies, often referred to as ``warm-starts''~\cite{Puig_2025}, is to place the initial parameters in a more favorable region of the optimization landscape: one with significant gradients and proximity to a good solution.

Several such strategies, drawing inspiration from both classical machine learning and established practices in quantum chemistry, have been proposed and have shown empirical success. 
These include:

\begin{itemize}
    \item \textbf{Restricted small angle initializations}, where parameters are still chosen randomly but from a constrained, structured region, such as near zero to approximate the identity transformation~\cite{grant2019initialization,jain2022graph,skolik2021layerwise,kulshrestha2022beinit,rad2022surviving,astrakhantsev2023phenomenological,haug2021optimal,kashif2024alleviating,shi2024avoiding}.
    \item \textbf{Pre-training}, where parameters are first optimized using classical~\cite{grimsley2019adaptive,friedrich2022avoiding,rudolph2023synergistic,marin2023quantum} or smaller, tractable quantum methods~\cite{cervera2021meta,goh2023lie}.
    \item \textbf{Parameter transfer}, which leverages solutions from smaller problem instances to construct an initial guess for larger, related ones~\cite{shaydulin2023parameter,farhi2022quantum,brandao2018fixed,zhou2020quantum,wurtz2021fixed,boulebnane2021predicting,galda2021transferability,mele2022avoiding,liu2023mitigating}.
\end{itemize}

The success and ongoing refinement of these methods underscore a key principle: the ability to incorporate problem-specific structure into the initial state is crucial for overcoming the challenge of barren plateaus. 
This sets the stage for our proposed method, which provides a systematic way to construct such a highly informed initial state.

\subsection{The EQS approach: combining inductive bias and a warm-start}
The mitigation strategies discussed in the previous sections highlight a clear path forward: overcoming barren plateaus requires moving beyond generic, randomly initialized circuits and instead embedding problem-specific structure into the variational algorithm. 
Our EQS approach provides a systematic framework to achieve this, uniquely combining a tailored ansatz architecture with a deterministic warm-start.

The EQS procedure achieves this through a three-pronged approach:
\begin{enumerate}
    \item \textbf{A targeted warm start}
    
    First and foremost, the EQS framework inherently provides an effective warm start. 
    Unlike other informed initializations that can be heuristic or stochastic, such as choosing restricted small angles~\cite{grant2019initialization,jain2022graph,skolik2021layerwise,kulshrestha2022beinit,rad2022surviving,astrakhantsev2023phenomenological,haug2021optimal,kashif2024alleviating,shi2024avoiding}, the EQS procedure is fully deterministic. 
    It does not provide a rough guess in a promising region of the landscape, but instead calculates the specific parameter values required to realize the high-quality solution derived from the trained kernel model to arbitrary precision.
    By placing the initial point of the optimization, by construction, at this known good solution, we entirely circumvent the inefficient search through the vast, flat regions that a random initialization would have to navigate.

    \item \textbf{Adaptive circuit construction}
    
    Beyond the warm start, a second layer of BP mitigation arises from our specific implementation of the circuit construction step. 
    Within the EQS framework, we employ the AQCE algorithm~\cite{Shirakawa2021-uj}, which has an inherent structural feature. 
    Rather than optimizing a fixed, deep ansatz, AQCE iteratively grows the circuit from a simple starting point, adaptively adding gates to increase complexity. 
    This process is analogous to variational structured ansatze, a strategy well-documented in the literature for its ability to navigate optimization landscapes and avoid barren plateaus~\cite{bilkis2023semi,du2022quantum,Larocca_2025}. 
    Thus, our choice of AQCE as the construction tool provides an additional, mechanism-based defense against trainability issues.

    \item \textbf{An embedded inductive bias}
    
    Finally, the EQS framework provides an inductive bias by defining a specific target for the circuit construction, which contrasts sharply with approaches that use a fixed, generic ansatz. 
    The core of this bias is the target of the construction process itself: the eigenvectors of the trained kernel observable. 
    These eigenvectors dictate a non-arbitrary structure for the final circuit, fundamentally tailoring it to represent a solution already known to be effective for the given data distribution. 
    This design principle constrains the vast search space to a more relevant and promising region. 
    It is the constructive, ansatz-free nature of the AQCE algorithm that makes it possible to translate this abstract inductive bias into a concrete, physical quantum circuit.
\end{enumerate}

In summary, EQS is not merely an initialization technique but a comprehensive approach that synergistically combines an adaptive construction process, an architecturally embedded inductive bias, and a deterministic warm start.
This three-pronged strategy directly addresses the limitations of purely heuristic methods by providing a principled, problem-specific starting point. 
By doing so, it systematically mitigates the risk of barren plateaus and enhances the overall trainability of the model.

\section{On the distinction between barren plateaus and kernel concentration}
\label{sec: relation}
A crucial distinction must be made between the barren plateaus (BPs)~\cite{Larocca_2025} affecting explicit models and the exponential concentration~\cite{thanasilp2024exponential} affecting kernel methods, especially concerning model trainability.
A barren plateau is fundamentally a failure of the optimization process. 
The non-convex landscape of a parameterized quantum circuit can exhibit exponentially vanishing gradients, which prevents a gradient-based optimizer from finding a good solution.
In contrast, for a quantum kernel method, the representer theorem~\cite{Smola1998-zt,Scholkopf2002-mh} ensures the optimal model's structure allows the optimization problem to be cast as a convex one. 
For such problems, a globally optimal solution that minimizes the empirical loss is always guaranteed. 
Therefore, the issue with kernel concentration is not a failure to train, but a failure to generalize; the model can always be found, but it may not be useful for new data.

The fundamental nature of this distinction is highlighted by recent work showing that kernel-based pre-training can make an otherwise untrainable variational circuit trainable~\cite{gilfuster2025relation}. 
This confirms that the challenge for kernels lies in generalization, not in the optimization process itself, for which a globally optimal solution is guaranteed.

\section{Justification for the learnable regime}
\label{sec:learnable_regime}
The EQS framework is predicated on the distinction between generalization and trainability challenges. 
The issue of generalization brings us to a fundamental principle in machine learning. 
The no free lunch (NFL) theorem~\cite{wolpert1996lack} indicates that successful learning requires an algorithm’s inductive bias to align with the data’s inherent structure. 
In kernel methods, this crucial inductive bias is introduced primarily through the choice of the kernel function~\cite{Scholkopf2002-mh, Hofmann_2008}.

Our framework adheres to this principle by intentionally scoping our work to what we term the ``learnable regime''. 
We define this as the class of structured problems where the inductive bias of the global fidelity quantum kernel is effective. 
This bias is the assumption that proximity in Hilbert space is a meaningful measure of similarity, and its effectiveness ensures that exponential concentration~\cite{thanasilp2024exponential} is not a prohibitive barrier to generalization.
In this regime, the solution found via convex optimization is both globally optimal for the training data and meaningful for unseen data.

The EQS framework then provides a deterministic bridge to carry this guaranteed, high-performance solution from the easy-to-solve convex world into the hard-to-search non-convex world of explicit models. 
In doing so, EQS initialization is expected to circumvent the inefficient search through barren plateau landscapes that randomly initialized models would otherwise face.

\section{Resource cost and scalability analysis}
\label{sec:cost_analysis} 
In this section, we provide a detailed analysis of the required computational costs and scalability of the EQS framework. 
We perform the analysis separately for two phases: the one-time construction of the EQS and its subsequent use for prediction. 
The construction requires an upfront investment of resources during the training phase. 
We show that the primary advantage of this investment is the improvement in prediction efficiency.

\subsection{EQS construction cost}
\subsubsection{Quantum cost (measurement shots)}
The most direct measure of the quantum resources consumed during the construction phase is the total number of Hadamard tests performed by the iterative AQCE algorithm~\cite{Shirakawa2021-uj}. 
A key challenge, however, is that because AQCE is a heuristic algorithm, this quantity cannot be determined a priori; it is an empirical value that is highly dependent on the specific problem instance.
A formal, theoretical cost analysis is therefore intractable at this stage. To provide a practical, empirical measure of the construction effort, our analysis instead uses the number of two-qubit gates in the final circuit as a proxy. 
This choice is justified by the fact that a more complex final circuit (i.e., a higher gate count) generally reflects a more extensive and costly optimization search.

With this in mind, we perform a numerical simulation to address a key question regarding this proxy cost under realistic conditions: How significantly does shot noise increase the required gate count for the EQS construction?

To answer this question, we conduct an end-to-end simulation of the construction process under two scenarios: a noiseless environment and a noisy one with a finite budget of $10^6$ shots per measurement. 
The simulation begins with a single, noiselessly pre-trained implicit model. 
For both scenarios, we then analyze the number of two-qubit gates required for the AQCE algorithm to reach a target fidelity of $F^{(k)}>0.6$ for each target eigenvector. 
Due to the high computational cost of the noisy simulation, we construct the EQS models using a reduced training set of $50$ samples per label ($500$ total) from the MNISQ-MNIST dataset~\cite{Placidi2023-kq}.

\begin{figure}
\centering
\includegraphics[width=.6\linewidth]{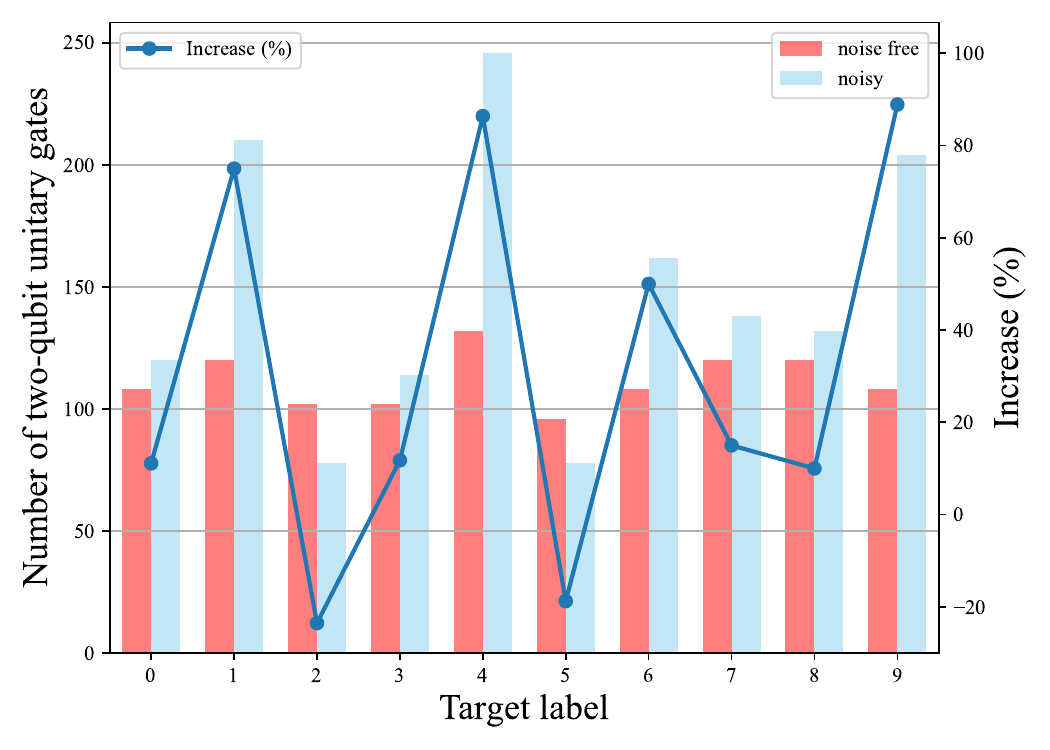} 
\caption{\textbf{Impact of shot noise on EQS construction cost.} 
The number of two-qubit gates required to construct the circuit for each label of the MNISQ-MNIST dataset, such that the fidelity for each eigenvector satisfies $F^{(k)}>0.6$, under noiseless (red bars) and noisy (blue bars, $10^6$ shots) conditions. 
The line plot shows the percentage increase in gate count due to shot noise.}
\label{fig:gate_overhead}
\end{figure}

The result of this simulation is presented in Fig.~\ref{fig:gate_overhead}. 
While shot noise increases the required number of two-qubit gates to reach the target fidelity, the gate count increases by less than 20\% for 6 out of 10 labels.
Even for the remaining labels that saw a more significant increase, this provides evidence that the optimization process remains tractable under realistic noise conditions. 
While $10^6$ shots per estimate represents a significant experimental cost, this level of sampling is typical in studies aiming to suppress statistical noise to a level where it does not obscure the underlying algorithmic performance.

\subsubsection{Classical cost}
The classical computation is dominated by the diagonalization of the trained observable $O_{\bm{\alpha},\mathcal{D}}$ from the implicit model.
In the worst case, this requires classical resources scaling as $O(M^3)$, where $M$ is the number of training data points. 
We note that this cost is comparable to the classical cost of standard kernel methods, which often require inverting an $M \times M$ kernel matrix, also an $O(M^3)$ operation.
While an $O(M^3)$ scaling can be demanding, this cost is manageable for moderately large $M$ on modern high-performance computing systems.
Therefore, the classical computational cost is not expected to be the overall bottleneck in many cases compared to the quantum resources required for EQS construction.

\subsection{Prediction cost}
\label{sec: prediction_cost}
Once the EQS is constructed, the prediction for a new data point $\boldsymbol{x}$ is obtained by evaluating $f_{\mathrm{EQS}}(\boldsymbol{x}) = \mathrm{Tr}[O\mathcal{C}^\dagger\rho(\boldsymbol{x})\mathcal{C}]$. 
Because the observable $O=\sum_{k=0}^{K-1}\lambda_{k}|k\rangle \langle k|$ is diagonal in the computational basis, this expectation value can be estimated by preparing the state $|\psi_{\mathrm {out }}\rangle = \mathcal{C}^\dagger U(\boldsymbol{x})|0\rangle$ and measuring it in the computational basis just once to sample from the probability distribution $p(k|\boldsymbol{x}) = |\langle k | \psi_{\mathrm {out }} \rangle|^2$.
Therefore, the quantum cost in terms of the number of distinct circuit executions is $O(1)$, a sharp contrast to the $O(M)$ cost of the original implicit model. 

Furthermore, the classical computation required to obtain the final prediction value from the measurement outcomes involves only negligible arithmetic operations.
However, a potential concern is the statistical cost.
Specifically, one might worry that if the state $|\psi_{\mathrm{out }}\rangle$ is highly delocalized (e.g., similar to a Haar-random state), the probability of measuring the relevant outcomes $\{|k\rangle\}_{k=0}^{K-1}$ could be exponentially small, potentially requiring an exponential number of shots.
However, we now show that this is not the case based on Appendix B.1 in~\cite{Jerbi2023-wt}.
This problem of estimating the expectation value is equivalent to a standard monte carlo mean estimation task. 
The measurement of the observable $O = \sum_{k=0}^{K-1} \lambda_k |k\rangle\langle k|$ yields outcomes corresponding to its eigenvalues $\{\lambda_k\}$, which are bounded within the interval $[-\lambda_{\max}, \lambda_{\max}]$, where $\lambda_{\max} = \|O\|_\infty$ is the spectral norm. 
It is a standard result from classical estimation theory that estimating the mean of a random variable bounded in $[-B, B]$ to a precision $\varepsilon$ with confidence $1-\delta$ requires a number of samples $N$ given by~\cite{dagum2000optimal,canetti1995lower}:
\begin{align}
    N = \Theta\left(\frac{B^2}{\varepsilon^2}\log\frac{1}{\delta}\right).
\end{align}
By setting $B = \lambda_{\max}$, we find that the sample complexity for estimating $f_{\mathrm{EQS}}(\boldsymbol{x})$ is efficient and independent of the system size. Crucially, this bound depends only on a property of the trained model, not on the state $|\psi_{\mathrm{out}}\rangle$ being measured.

\begin{figure}
\centering
\includegraphics[width=.4\linewidth]{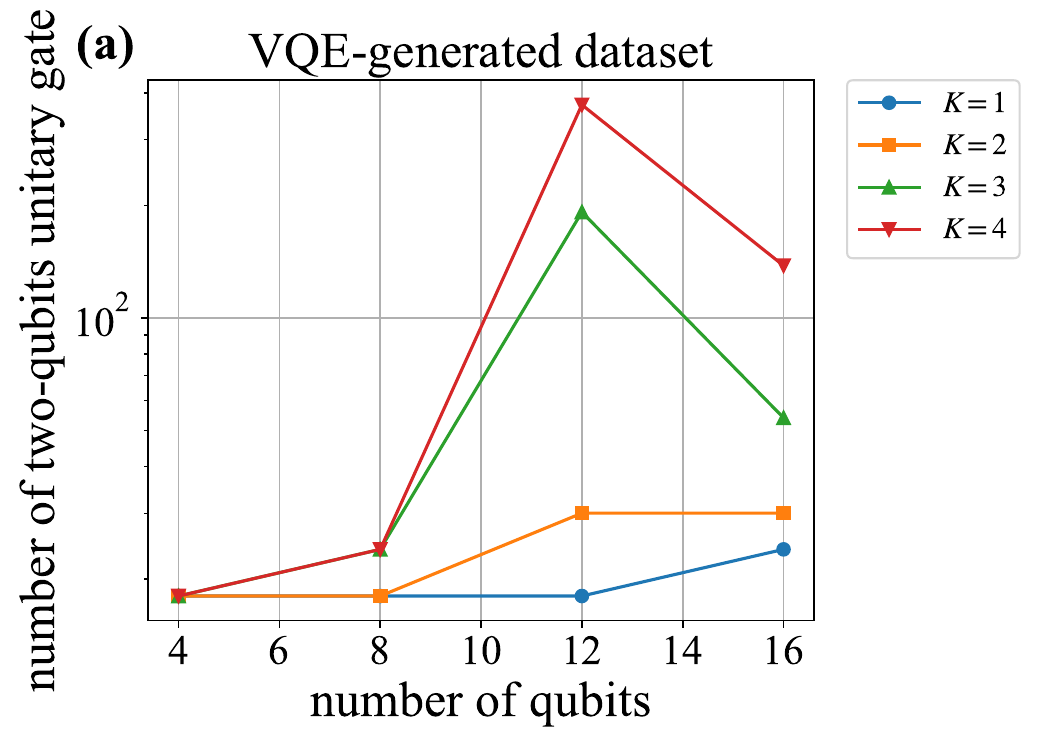}
\includegraphics[width=.4\linewidth]{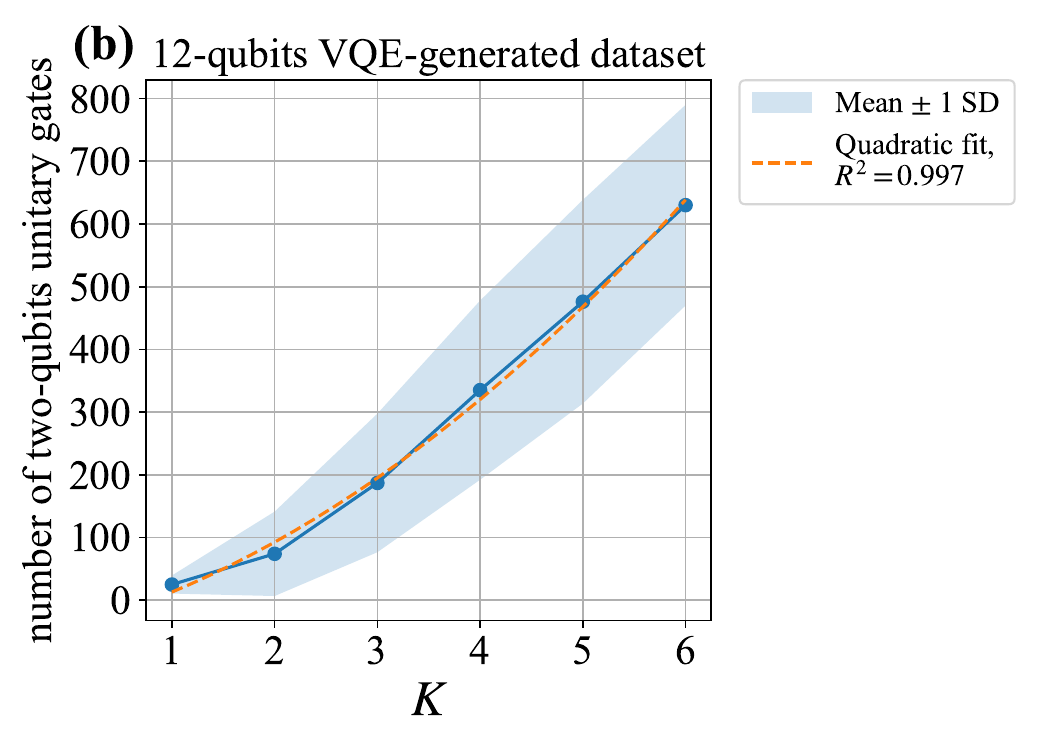}
\includegraphics[width=.4\linewidth]{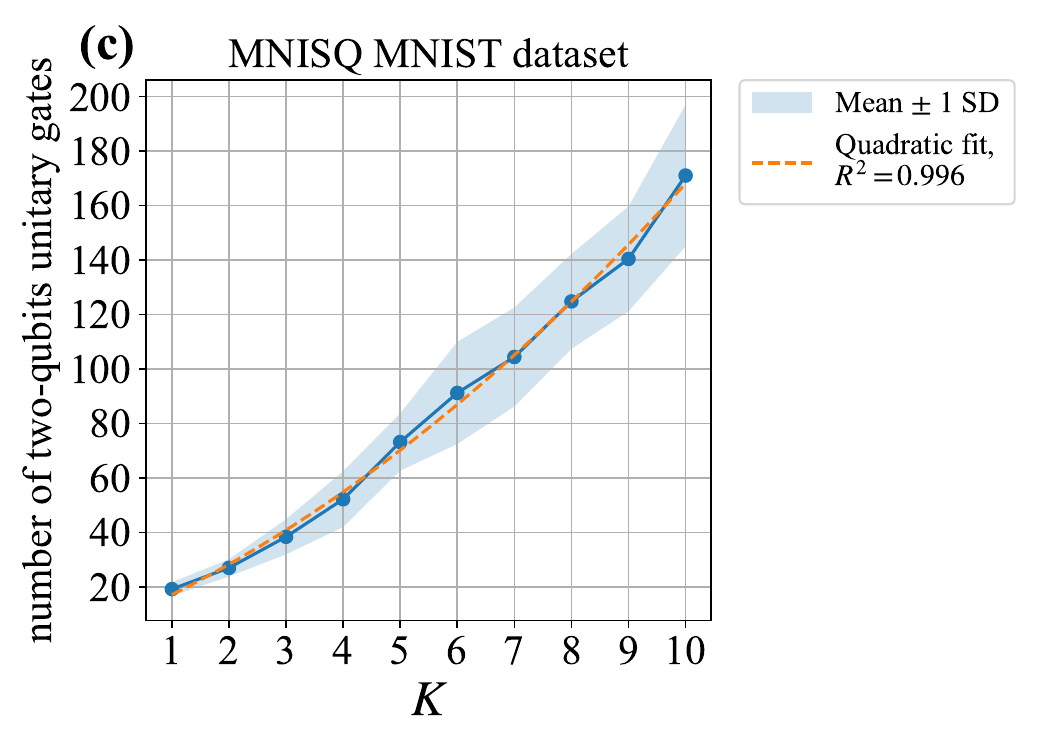}
\caption{Scalability analysis of the EQS circuit depth. 
(a) Scaling with the number of qubits $n$ for the VQE-generated dataset. 
The plot shows the result for a target label of 3. 
(b) Scaling with the number of embedded eigenvectors $K$ for the VQE-generated dataset.
(c) Scaling with the number of embedded eigenvectors $K$ for the MNISQ MNIST dataset.
In panels (b) and (c), the solid line represents the mean number of two-qubit gates averaged over all target labels, and the shaded area indicates the standard deviation.
}
\label{fig:appendix_scaling}
\end{figure}

\subsection{Scalability of the circuit construction}
A crucial question for the feasibility of the EQS framework is the scalability of its construction process. 
This is primarily determined by the number of two-qubit gates required by the circuit construction algorithm. 
While preparing an arbitrary, unstructured (e.g., Haar-random) state requires resources that scale exponentially with the number of qubits $n$, our framework operates within the ``learnable regime'' established in Appendix~\ref{sec:learnable_regime}. 
In this regime, the kernel captures the data's inherent structure, and the resulting eigenvectors--being linear combinations of these structured feature states--inherit this structure. 
This implies that the resources required to prepare them should not scale exponentially with the system size, an expectation we now investigate empirically.

To analyze the empirical scalability of the circuit construction step, we quantify the required number of two-qubit gates as a function of the number of qubits $n$ and the number of embedded eigenvectors $K$.
We use our extended AQCE algorithm on the VQE-generated datasets~\cite{Nakayama2023-zc} and MNISQ-MNIST datasets~\cite{Placidi2023-kq} under the noiseless simulation conditions detailed in Appendix~\ref{sec: condition}.

The results of this analysis are presented in Fig.~\ref{fig:appendix_scaling}.
First, regarding the scaling with $n$ (Fig.~\ref{fig:appendix_scaling}~(a)), the gate count shows no exponential growth within the tested range.
We believe this observation is linked to the intrinsic properties of the VQE-generated dataset itself.
The source quantum circuits within this dataset, each encoding an approximate ground state, were constructed with a fixed gate count independent of the number of qubits.
Since the eigenvectors of our EQS are linear combinations of states from these circuits, it is plausible that their construction complexity does not necessarily grow with the system size.
This result highlights that the scalability of the circuit construction cost is highly data-dependent and suggests that for datasets with an underlying learnable structure, the resources for constructing an EQS may not face exponential growth.
In contrast, scaling with the number of eigenvectors $K$ reveals a consistent property across both the VQE-generated (Fig.~\ref{fig:appendix_scaling}~(b)) and MNISQ-MNIST (Fig.~\ref{fig:appendix_scaling}~(c)) datasets. 
In both cases, the required number of two-qubit gates exhibits a quadratic trend with $K$.
This property has a practical advantage: it implies that we can increase the model's complexity and expressiveness by increasing $K$ without causing a prohibitive (e.g., exponential) increase in circuit depth.
This quadratic scaling, combined with our finding that high accuracy is achievable with small $K$ (see Fig.~\ref{fig:sim_result} in the main text), suggests that practical EQS models can be constructed with circuits of moderate depths.

\section{Automatic quantum circuit encoding}
\label{sec: aqce}
In this section, we briefly summarize the automatic quantum circuit encoding algorithm (AQCE)~\cite{Shirakawa2021-uj}.
AQCE is an algorithm that produces a quantum circuit $\mathcal{C}$, which outputs a quantum state $\mathcal{C}|0\rangle$ equivalent to a given arbitrary quantum state $|\Psi\rangle$ with the accuracy desired by the user.
AQCE sequentially updates the unitary gates that construct $\mathcal{C}$ using a gradient-free method to maximize the fidelity function $F = \left|\left\langle 0 \left| \mathcal{C}^{\dagger} \right| \Psi \right\rangle\right|$.

Below, we describe the AQCE algorithm in detail.
Assuming that the quantum circuit $\mathcal{C}$ is composed of $J$ two-qubit unitary gates $\mathcal{U}_m$ with $1\leq m \leq J$, namely,
\begin{equation}
    \mathcal{C}=\prod_{m=1}^J \mathcal{U}_m=\mathcal{U}_1 \mathcal{U}_2 \ldots \mathcal{U}_J.
    \label{eq:c}
\end{equation}
When considering an update for the $m$th unitary gate $\mathcal{U}_m$, it is convenient to define a fidelity function, 
\begin{align}
F_m=\left|\operatorname{Tr}\left[\left|\Psi_{m+1}\right\rangle\left\langle\Phi_{m-1}\right| \mathcal{U}_m^{\dagger}\right]\right|
\end{align}
that explicitly focuses only on the degrees of freedom of $\mathcal{U}_{m}$, where $\left|\Psi_m\right\rangle$ and $\left\langle\Phi_m\right|$ are defined by
\begin{align}
    \left|\Psi_m\right\rangle
    &=\prod_{j=m}^J \mathcal{U}_j^{\dagger}|\Psi\rangle=\mathcal{U}_m^{\dagger} \mathcal{U}_{m+1}^{\dagger} \ldots \mathcal{U}_J^{\dagger}|\Psi\rangle,\\
    \left\langle\Phi_m\right|
    &=\langle 0| \prod_{j=1}^m \mathcal{U}_j^{\dagger}=\langle 0| \mathcal{U}_1^{\dagger} \mathcal{U}_2^{\dagger} \ldots \mathcal{U}_m^{\dagger}.
\end{align}
If $\mathbb{I}_m$ denotes the set of indices $\{i, j\}$ corresponding to the qubits on which the unitary gate $\mathcal{U}_m$ acts, then $F_m$ can be rewritten as
\begin{align}F_m=\left|\operatorname{Tr}_{\mathbb{I}_m}\left[\mathcal{F}_m \mathcal{U}_m^{\dagger}\right]\right|,
\end{align}
where $\mathcal{F}_m$ is referred to as the fidelity tensor operator and is defined by the equation
\begin{equation}
\label{eq:fidelity_tensor_original}
\mathcal{F}_m=\operatorname{Tr}_{\bar{\mathbb{I}}_m}\left[\left|\Psi_{m+1}\right\rangle\left\langle\Phi_{m-1}\right|\right].
\end{equation}
Here, $\bar{\mathbb{I}}_m$ is the complement of the subsystem $\mathbb{I}_{m}$ in the total qubit system.
If we represent the fidelity tensor operator $\mathcal{F}_m$ and the unitary gate $\mathcal{U}_m$ in matrix form as $\boldsymbol{F}_{m}$ and $\boldsymbol{U}_{m}$, respectively, the expression is given by
\begin{equation}
\label{eq:aqce_form}
F_m = \left|\operatorname{Tr}\left[\boldsymbol{F}_m \boldsymbol{U}_m^\dagger\right]\right|.
\end{equation}

The AQCE algorithm updates the unitary gates $\mathcal{U}_m$ to maximize the fidelity function $F_m$.
This can be achieved through singular value decomposition of $\boldsymbol{F}_{m}$, expressed as $\bm{F}_m=\boldsymbol{X D Y}$, where $\boldsymbol{X}$ and $\boldsymbol{Y}$ are unitary matrices, and $\boldsymbol{D}$ is a diagonal matrix with non-negative diagonal elements $d_n$.
The fidelity function $F_m$ is then given by
\begin{align}    
F_{m}&=\left|\operatorname{Tr}\left[\boldsymbol{X} \boldsymbol{D} \boldsymbol{Y} \boldsymbol{U}_{m}^{\dagger}\right]\right|\\
&=\left|\operatorname{Tr}[\boldsymbol{D} \boldsymbol{Z}]\right|\\
&=\left|\sum_{n=0}^{3} d_{n}[\boldsymbol{Z}]_{n n}\right|\\
&\leq \sum_{n=0}^3 d_n\left|[\boldsymbol{Z}]_{n n}\right|,
\end{align}
where we defined a unitary matrix $\boldsymbol{Z}=\boldsymbol{Y} \boldsymbol{U}_{m}^{\dagger} \boldsymbol{X}$.
From this, $\boldsymbol{Z}$, which maximizes the fidelity function $F_m$, satisfies $|[\boldsymbol{Z}]_{n n^{\prime}}|=\delta_{n n^{\prime}}$.
Consequently, a unitary gate $\mathcal{U}_m$ that maximizes the fidelity function $F_m$ is given by
\begin{equation}
\bm{U}_m=\bm{XY}.
\end{equation}

The pseudocode for the AQCE algorithm is presented in Alg.~\ref{algo:aqce}.
While the process in line 7 is computationally expensive on classical computers, it can be efficiently computed using a quantum computer with the Hadamard test~\cite{Shirakawa2021-uj}.
In contrast, the calculations from lines 8 to 10 can be efficiently performed on a classical computer.

\begin{algorithm}[H]
\caption{AQCE}
\label{algo:aqce}
\textbf{Inputs:} Quantum state $|\Psi\rangle$, initial number of unitary gates $J_0\in \mathbb{N}$, increase in the number of unitary gates per sweep $\delta J\in \mathbb{N}$, number of sweep $N\in \mathbb{N}$, maximum number of unitary gates $J_{\max}\in \mathbb{N}$, target fidelity $F_{\mathrm{target}}\in (0,1]$, set of indices where a two-qubit unitary can be placed $\mathbb{B}$ \\
\textbf{Output:} Quantum circuit $\mathcal{C}$ 	
 \begin{algorithmic}[1]
        \State Initialization: $\mathcal{C} \leftarrow I \text{ and } J \leftarrow J_0$
        \While{$J < J_{\max} \wedge F < F_{\mathrm{target}}$}
		  \State Add $\delta J$ two-qubit gates to $\mathcal{C}$
          \State $J \leftarrow J + \delta J$
          \For{$\mathrm{counter} = 1$ to $N$}
            \For{$m = 1$ to $J$}
              \State For all indices $\mathbb{I}_l \in \mathbb{B}$, find the representation matrix $\bm{F}_{m,l}$ for $\mathcal{F}_{m,l} = \operatorname{Tr}_{\bar{\mathbb{I}}_l}\left[\left|\Psi_{m+1}\right\rangle\left\langle\Phi_{m-1}\right| \right]$.
              \State For all $\bm{F}_{m,l}$, perform the singular value decomposition $\boldsymbol{F}_{m,l} = \boldsymbol{X}_l \boldsymbol{D}_l \boldsymbol{Y}_l$ and compute $S_l = \sum_{n=0}^{3}\left[\boldsymbol{D}_l\right]_{nn}$
              \State Find $l = l^*$ that maximizes $S_l$.
              \State Calculate $\boldsymbol{U}_m^* = \boldsymbol{X}_{l^*} \boldsymbol{Y}_{l^*}$ and determine the unitary gate $\mathcal{U}^*_{m,l^*}$ that corresponds to $\boldsymbol{U}^*_m$ and acts on $\mathbb{I}_{l^*}$.
              \State $\mathcal{U}_m \leftarrow \mathcal{U}^*_{m,l^*}$
            \EndFor
          \EndFor
		\EndWhile
	\end{algorithmic}
\end{algorithm}

\section{Extending AQCE to isometries}
\label{sec: extended_aqce}
In this section, we detail the extension of AQCE to isometries.
More concretely, our extended AQCE is an algorithm designed to generate a quantum circuit $\mathcal{C}$ that satisfies $\mathcal{C}|k\rangle\approx|\Psi^{(k)}\rangle$ for a set of orthogonal quantum states $\{|\Psi^{(k)}\rangle\}_{0\leq k \leq K-1}$, under the condition that the global phase of quantum states is disregarded.
Similar to AQCE, we assume that the quantum circuit is composed of $J$ two-qubit unitary gates, as shown in Eq.~(\ref{eq:c}). 
We modify the fidelity functions $F$ and $F_m$ from AQCE as follows:
\begin{align}
F &=\sum_{k=0}^{K-1} F^{(k)}=\sum_{k=0}^{K-1} \left|\left\langle k\left|\mathcal{C}^{\dagger}\right| \Psi^{(k)}\right\rangle\right|,
\end{align}
For convenience, we define 
\begin{align}
F_m=\sum_{k=0}^{K-1}\left|\operatorname{Tr}\left[\left|\Psi_{m+1}^{(k)}\right\rangle\left\langle\Phi_{m-1}^{(k)}\right| \mathcal{U}_m^{\dagger}\right]\right|,
\end{align}
where
\begin{align}
    \left|\Psi_m^{(k)}\right\rangle&=\prod_{j=m}^J \mathcal{U}_j^{\dagger}|\Psi^{(k)}\rangle
=\mathcal{U}_m^{\dagger} \mathcal{U}_{m+1}^{\dagger} \ldots \mathcal{U}_J^{\dagger}|\Psi^{(k)}\rangle,\\
    \left\langle\Phi_m^{(k)}\right|&=\langle k| \prod_{j=1}^m \mathcal{U}_j^{\dagger}=\langle k| \mathcal{U}_1^{\dagger} \mathcal{U}_2^{\dagger} \ldots \mathcal{U}_m^{\dagger}.
\end{align}
Using $\mathbb{I}_m$ defined in Appendix~\ref{sec: aqce}, $F_m$ can be rewritten as
\begin{align}
    F_m=\sum_{k=0}^{K-1}\left|\operatorname{Tr}_{\mathbb{I}_m}\left[\mathcal{F}_m^{(k)} \mathcal{U}_m^{\dagger}\right]\right|.
\end{align}
Here, $\mathcal{F}_m^{(k)}$ is defined as:
\begin{equation}
\label{eq:fidelity_tensor}
\mathcal{F}_m^{(k)}=\operatorname{Tr}_{\bar{\mathbb{I}}_m}\left[\left|\Psi_{m+1}^{(k)}\right\rangle\left\langle\Phi_{m-1}^{(k)}\right|\right].
\end{equation}
If we represent the fidelity tensor operator $\mathcal{F}_m^{(k)}$ and the unitary gate $\mathcal{U}_m$ as matrices $\boldsymbol{F}_{m}^{(k)}$ and $\boldsymbol{U}_{m}$, respectively, we can reformulate the fidelity function $F_m$ as follows:
\begin{equation}
F_m=\sum_{k=0}^{K-1}\left|\operatorname{Tr}\left[\boldsymbol{F}_m^{(k)} \boldsymbol{U}_m^\dagger\right]\right|.
\end{equation}

To optimize the unitary gate $\mathcal{U}_m$ to maximize the fidelity function $F_m$ using the AQCE algorithm strategy, we reformulate the fidelity function $F_m$ in the same form as in Eq.~\eqref{eq:aqce_form}.
This can be achieved if 
 $\operatorname{Tr}\left[\boldsymbol{F}_m^{(k)} \boldsymbol{U}_m^\dagger\right]$ is always real and positive in all cases.
To achieve this, we transform $|\Psi^{(k)}\rangle$ as follows to cancel its global phase:
\begin{align}
    |\Psi^{(k)}\rangle \rightarrow e^{i\theta^{(k)}}|\Psi^{(k)}\rangle,
\end{align}
where $\theta^{(k)} \in [0,2\pi)$.
As a result, $F_m$ can be written as
\begin{align}
    F_{m}
    &=\sum_{k=0}^{K-1}\left|\operatorname{Tr}\left[e^{i\theta_m^{(k)}}\boldsymbol{F}_m^{(k)} \boldsymbol{U}_m^\dagger\right]\right|=\left|\operatorname{Tr}\left[\boldsymbol{F}_{m} \boldsymbol{U}_m^\dagger\right]\right|,
\end{align}
where $\boldsymbol{F}_{m}=\sum_{k=0}^{K-1}e^{i\theta_m^{(k)}}\boldsymbol{F}_m^{(k)}$.
It should be noted that changes in the global phase of quantum states $|\Psi^{(k)}\rangle$ do not affect the construction of our EQS.
This is because low-rank approximations of an observable of implicit models and the fidelity function $F_m$ are invariant with respect to the global phase of quantum states $|\Psi^{(k)}\rangle$. 
Furthermore, it is worth mentioning that even if some applications require the global phase, simply applying this phase to the initial state before applying the quantum circuit suffices.
In this case, we update the unitary gate $\mathcal{U}_m$ following the update of the phase $\theta_m^{(k)}$.
The optimization cost of the phase $\theta_m^{(k)}$ is almost negligible compared to the fidelity tensor in Eq.~\eqref{eq:fidelity_tensor}.
We present the pseudo-algorithm for the extended AQCE in Alg.~\ref{algo:extended_aqce}.

We note that it is possible to update the unitary gate $\mathcal{U}_m$ to maximize the fidelity function using gradient methods, without granting additional degrees of freedom to the phase of the quantum state $|\Psi^{(k)}\rangle$.
Specifically, we parameterize the matrix representation of the two-qubit unitary gate $\mathcal{U}_m$, denoted as $\boldsymbol{U}_m$, as $\boldsymbol{U}_m(\boldsymbol{\theta}_m)$.
We then optimize parameters $\boldsymbol{\theta}_m$ to maximize the fidelity function $F_m=\sum_{k=0}^{K-1} 
    \left|\left\langle\Phi_{m-1}^{(k)}\left|\boldsymbol{U}_m^{\dagger}(\boldsymbol{\theta}_m)\right| \Psi_{m+1}^{(k)}\right\rangle\right|$.
In this case, we do not need to compute the fidelity tensor operator, although we incur an additional cost in computing the gradient.
Therefore, it may be beneficial if the cost of computing the fidelity tensor operator exceeds the cost of computing the gradient.
Such situations may arise when using actual quantum computers for execution.

\begin{algorithm}[H]
    \caption{Extended AQCE}
	\label{algo:extended_aqce}
	\begin{algorithmic}[1]
		\State \textbf{Inputs:}
            Set of orthogonal quantum states $\{|\Psi^{(k)}\rangle\}_k$, 
		initial number of unitary gates $J_0\in \mathbb{N}$, 
		increase in the number of unitary gates per sweep $\delta J\in \mathbb{N}$, 
            number of sweep $N\in \mathbb{N}$, 		
            maximum number of unitary gates $J_{\max}\in \mathbb{N}$, 
            set of target fidelities $\left\{F_{\mathrm{target}}^{(k)}\right\}_k$, $F_{\mathrm{target}}^{(k)}\in (0,1]$ for the quantum state $|\Psi^{(k)}\rangle$, 
            set of indices where a two-qubit unitary can be placed $\mathbb{B}$ 
        \State \textbf{Output:} Quantum Circuit $\mathcal{C}$
        \State Initialization: $\mathcal{C} \leftarrow I \text { and } J \leftarrow J_0$
		  \While{$J< J_{\max} \ \wedge\ \min_k\,F^{(k)} < F^{(k)}_{\mathrm{target}}$}
          \State Add $\delta J$ two-qubit gates to $\mathcal{C}$
          \State $J \leftarrow J + \delta J$
          \For{$\mathrm{counter} = 1$ to $N$}
            \For{$m = 1$ to $J$}
              \State For all indices $\mathbb{I}_l \in \mathbb{B}$ and $k$, find the representation matrix $\bm{F}_{m,l}^{(k)}$ for $\mathcal{F}_{m,l}^{(k)}=\operatorname{Tr}_{\bar{\mathbb{I}}_l}\left[\left|\Psi_{m+1}^{(k)}\right\rangle\left\langle\Phi_{m-1}^{(k)}\right|\right]$.
              \State For all indices $\mathbb{I}_l\in\mathbb{B}$ and $k$, find the phase component $\phi_{m,l}^{(k)}$ for 
              $\operatorname{Tr}\left[ \bm{F}_{m,l}^{(k)} \bm{U}^\dagger_m\right]$.
              \State $\theta_{m,l}^{(k)} \leftarrow -\phi_{m,l}^{(k)}$ for all $l$ and $k$
              \State Calculate $\boldsymbol{F}_{m,l}=\sum_{k=0}^{K-1}e^{i\theta_{m,l}^{(k)}}\boldsymbol{F}_{m,l}^{(k)}$ for all $l$
              \State Execute steps 8-11 in Alg.~\ref{algo:aqce}.
            \EndFor
          \EndFor
		\EndWhile
	\end{algorithmic}
\end{algorithm}

\section{Detailed conditions of numerical experiments}
\subsection{Dataset descriptions}
\label{sec: dataset_descriptions}
Here we provide further details on the datasets used in our numerical experiments. 
For the tasks we consider, each input data point $\boldsymbol{x}$ is itself a quantum circuit, provided as a character string in QASM format~\cite{Cross2022-pb}. 
Consequently, the feature encoding map $U(\boldsymbol{x})$ is simply the execution of the circuit defined by that string. 
We selected two datasets with distinct structural properties to test our framework under different conditions.

\subsubsection{MNISQ dataset}
The MNISQ dataset~\cite{Placidi2023-kq} consists of 10-qubit quantum circuits that encode classical image vectors from the MNIST handwritten digit dataset~\cite{Lecun1998-dr}. 
The circuits were originally generated using the automatic quantum circuit encoding (AQCE) algorithm~\cite{Shirakawa2021-uj}, which constructed each circuit $U$ such that its output state $U|0\rangle\langle0|U^\dagger$ represents the corresponding classical image vector. 
A key property of this dataset is that circuits with the same digit label are designed to produce similar quantum states, representing a realistic scenario where intra-class similarity is high.

\subsubsection{VQE-generated dataset}
The VQE-generated dataset~\cite{Nakayama2023-zc} is composed of 4- to 20-qubit quantum circuits obtained from executing the variational quantum eigensolver (VQE) algorithm~\cite{Peruzzo2014-gz,Tilly2022-nh}. 
In contrast to the MNISQ dataset, this dataset is derived from a quantum-native problem: the classification of physically meaningful quantum states. 
Each circuit prepares an approximate ground state for a specific physical Hamiltonian (e.g., the transverse-field Ising model).
A defining characteristic of this dataset is that states corresponding to different classes (i.e., ground states of different hamiltonians) are known to be nearly orthogonal. 
This provides a distinct test case for our framework on a quantum-native classification task.

\subsection{SVM and EQS construction details}
\label{sec: condition}
First, we provide the details for training the implicit SVM model and constructing the EQS models for the numerical experiments presented in Fig.~\ref{fig:sim_result}.
As our implicit model, we employ a kernel support vector machine (SVM) model~\cite{Pl1999,Wu2003-nv}.
We employ the one-vs-rest approach for multi-class classification; that is, we train a separate SVM model $f_{\mathrm{implicit}}^{(l)}$ for each class $l$ to distinguish it from all other classes.
The overall prediction for an input $\bm{x}$ is made by outputting $\mathrm{argmax}_l f_{\mathrm{implicit}}^{(l)}(\bm{x})$.
The SVM model is trained using scikit-learn 1.3.0~\cite{scikit-learn} with a regularization strength of $C=1.0$.
To construct EQS, we first perform an exact diagonalization of $O_{\boldsymbol{\alpha}, \mathcal{D}}^{(l)}$ to determine its eigenvalues $\{\lambda_i^{(l)}\}_i$ and eigenvectors $\{\ket{\lambda_i^{(l)}}\}_i$.
Then, we generate quantum circuits $\mathcal{C}^{(l)}$ using Alg.~\ref{algo:extended_aqce}.
The parameters for the algorithm are set as $F_{\mathrm {target }}^{(0)}=0.6, \cdots, F_{\mathrm {target }}^{(K-1)}=0.6$, $J_0=12$ and $\delta J=6$, without specifying $J_{\max}$.
We run noiseless quantum circuit simulations using Qulacs 0.5.6~\cite{Suzuki2021-nj}.
For the input data, we focus on the MNIST dataset from the MNISQ dataset~\cite{Placidi2023-kq}, which has a fidelity of over 95\%.
We sample 1,000 data points for each label from the dataset, totaling 10,000 data points across all labels.
Half of the data from each label is allocated for training, with the remaining half designated for testing.
Additionally, we focus on 12-qubit VQE-generated datasets~\cite{Nakayama2023-zc}.
The datasets comprise six labels.
For each label, we use all 300 data points, resulting in a total of 1,800 data points.
Half of the data for each label is allocated for training, with the remaining half used for testing.

Next, building on the base EQS models constructed as described above, we now provide the specific details for the numerical experiments in Fig.~\ref{fig:grad_result}.
These experiments focus on the VQE-generated datasets to perform a scaling analysis with the number of qubits $n$.
The core of our experiment is to compare the initial loss gradients under different initialization schemes for each target label $l$.
To do this, we first take the circuit ansatz $\mathcal{C}^{(l)}$ found by our extended AQCE algorithm. 
We then treat it as a parameterized quantum circuit (PQC), denoted $\mathcal{C}^{(l)}(\bm{\theta})$, by considering each of its two-qubit unitary gates as an arbitrary two-qubit unitary with 15 independent parameters.
We then precisely identify the advantage of our strategy by comparing the gradients under three distinct initialization schemes for the parameters $\bm{\theta}$:
\begin{enumerate}
    \item \textbf{EQS-initialized model:} 
    This is our proposed method. 
    We set the parameters $\bm{\theta}$ to the specific values originally determined by our extended AQCE algorithm.

    \item \textbf{EQS ansatz with random parameters:}
    As a control to test the benefit of the circuit structure alone, we use the same PQC ansatz $\mathcal{C}^{(l)}(\bm{\theta})$ but initialize its parameters $\bm{\theta}$ randomly from the interval $[0, 2\pi)$. 

    \item \textbf{Random ansatz with random parameters:} 
    As a baseline, we construct a PQC ansatz of the same gate count as that of $\mathcal{C}^{(l)}(\bm{\theta})$, but with randomly chosen gate positions. 
    Its parameters are also initialized randomly from $[0, 2\pi)$.
\end{enumerate}

To calculate the gradients for these models, we first define the components leading to our loss function.
First, the raw output of the model for label $l$ is given by:
\begin{align}\label{eq:eqs-induced-model}
f^{(l)}(\boldsymbol{x};\bm{\theta}) = \mathrm{Tr}[O^{(l)}\mathcal{C}^{(l)}(\bm{\theta})^\dagger \rho(\bm{x})\mathcal{C}^{(l)}(\bm{\theta})],
\end{align}
where $O^{(l)}=\sum_{k=0}^{K-1}\lambda_{k}^{(l)}|k\rangle\langle k|$.
Next, this output is mapped to the interval $(0, 1)$ using the sigmoid function, yielding a value $p^{(l)}$:
\begin{align}
p^{(l)}(\bm{x};\bm{\theta})=\frac{1}{1+e^{-f^{(l)}(\bm{x};\boldsymbol{\theta})}}.
\end{align}
Finally, using this value, we define the weighted cross-entropy function as the loss on an input dataset $\mathcal{X}$:
\begin{align}\label{eq:cross-entropy}
    L^{(l)}(\bm{\theta}; \mathcal{X})=-\frac{1}{M} \sum_{\bm{x}\in \mathcal{X}}\left[
\frac{M_{\neq l}}{M} y_{\bm{x}} \log \left(p^{(l)}(\bm{x};\bm{\theta})\right) 
+\frac{M_l}{M} \left(1-y_{\bm{x}}\right) \log \left(1-p^{(l)}(\bm{x};\bm{\theta})\right) \right].
\end{align}
Here, $M=|\mathcal{X}|$, $M_l$ is the number of data points belonging to the label $l$, and $M_{\neq l}$ is the number of data points belonging to other labels, that is, $M_{\neq l}=\sum_{l'\neq l}M_{l'}$. 
$y_{\bm{x}} \in \{0,1\}$ is the label corresponding to the input data $\bm{x}$, where data belonging to $l$ are labeled as $1$, and all others are labeled as $0$.
The gradients of the loss function, $\frac{\partial L^{(l)}}{\partial \bm{\theta}}$, are computed using Qulacs 0.5.6~\cite{Suzuki2021-nj} and JAX 0.4.30 \cite{jax2018github}.
For $\mathcal{X}$, we sample $M_l = 150$ data points for each label from the dataset that were not used in the training of EQS, resulting in a total of $M=750$ data points for the 5-label datasets (4-qubit) and $M=900$ for the 6-label dataset (8-, 12-, and 16-qubit).
The quantum circuit simulation is performed under a noiseless environment.

\section{Experiments on additional datasets}
\label{sec: add_simu}
To demonstrate the generality of our findings in Fig.~\ref{fig:sim_result}, we repeated the same performance analysis on two additional datasets: the MNISQ-Fashion MNIST dataset~\cite{Placidi2023-kq} and an 8-qubit VQE-generated dataset~\cite{Nakayama2023-zc}. 
The simulation conditions were identical to those described in Appendix~\ref{sec: condition}, with only the datasets being replaced.

\begin{figure}
 \centering
 \includegraphics[width=.4\linewidth]{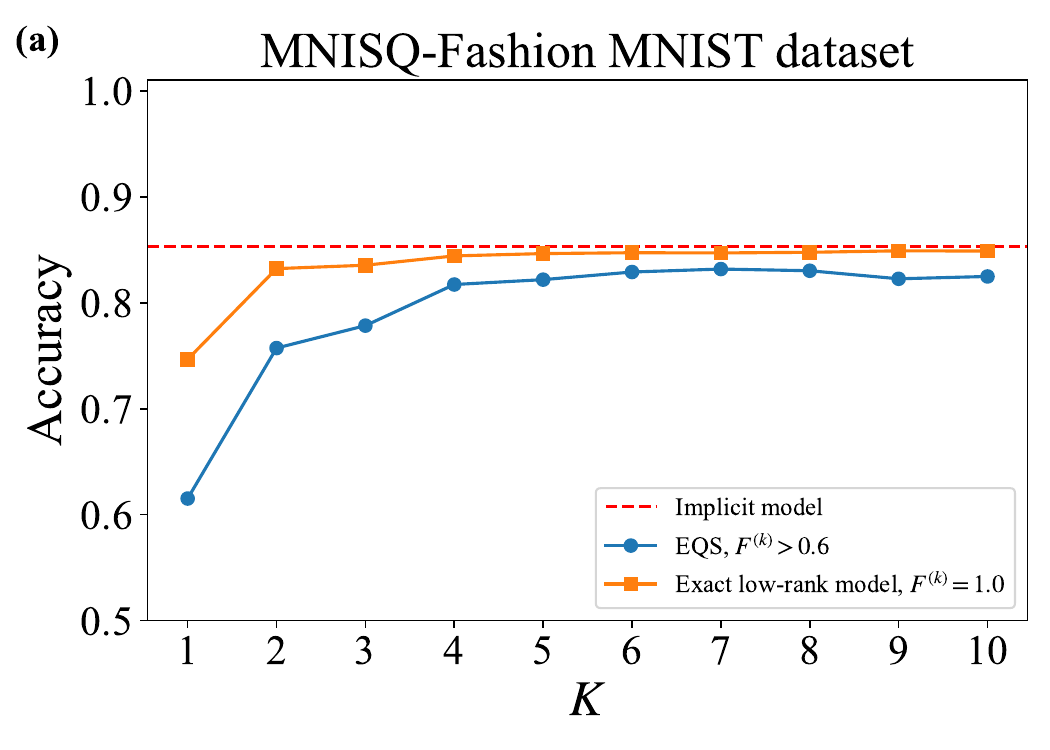}
 \includegraphics[width=.4\linewidth]{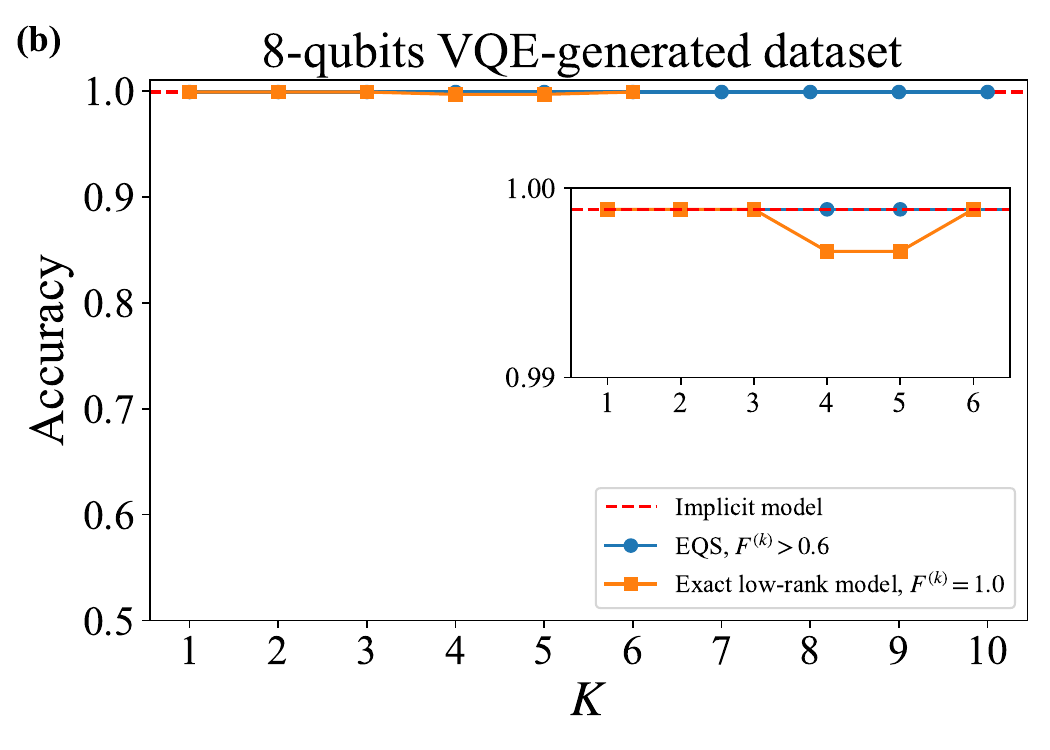}
 \caption{
 \textbf{Performance of EQS on MNISQ-Fashion MNIST and 8-qubit VQE-generated dataset.} 
 Notations follow that of Fig.~\ref{fig:sim_result} in the main text.
 }
 \label{fig:sup_svm_result}
\end{figure}

The results of these experiments, presented in Fig.~\ref{fig:sup_svm_result}, are consistent with those in the main text.
For the MNISQ-Fashion MNIST dataset (Fig.~\ref{fig:sup_svm_result}~(a)), the accuracy of the exact low-rank model improves and approaches the accuracy of the original implicit model as $K$ increases, showing only a 0.004 decrease compared to the full model at $K=10$.
In the case of the 8-qubit VQE-generated dataset (Fig.~\ref{fig:sup_svm_result}~(b)), the accuracy of the exact low-rank model saturates already at $K=1$.
This high accuracy is attributed to the unique structure of the VQE-generated dataset, as discussed in the main text. 
Since states with different labels are nearly orthogonal, classification can be effectively achieved using only the first eigenvector, which has high fidelity with states of the corresponding label, as shown in Fig.~\ref{fig:add_analysis_a} in Appendix~\ref{sec:additional_analysis}.

Furthermore, to confirm that the barren plateau mitigation effect shown in Fig.~\ref{fig:grad_result} is also a general phenomenon, we performed a gradient analysis on the 10-qubit MNISQ-MNIST dataset, following the same experimental protocol detailed in Appendix~\ref{sec: condition}.
As presented in Fig.~\ref{fig:grad_mnisq}, the results show a multi-order-of-magnitude gradient enhancement for the EQS-initialized model compared to randomly initialized models. 
This is consistent with the findings for the VQE-generated datasets~\cite{Nakayama2023-zc} discussed in the main text, and confirms that our initialization strategy may be effective across different data environments.

\begin{figure}
 \centering
 \includegraphics[width=.7\linewidth]{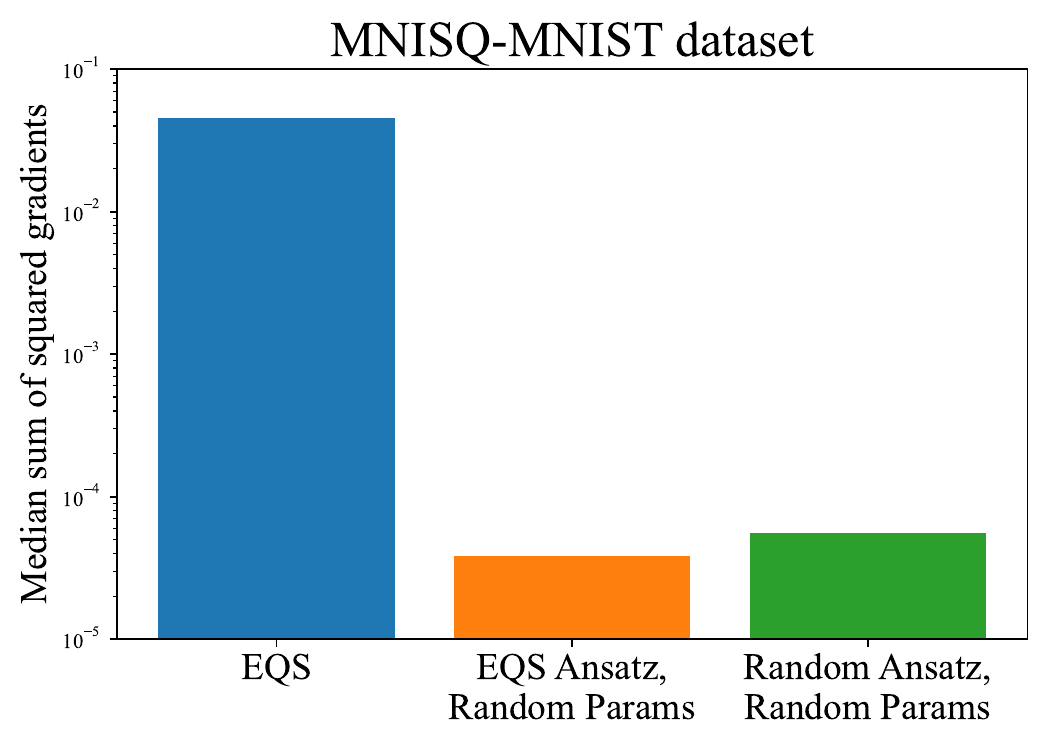} 
 \caption{
  \textbf{Sum of squared gradients for the MNISQ-MNIST dataset.} 
  This result shows that a multi-order-of-magnitude gradient enhancement from EQS initialization is a general phenomenon, not specific to the VQE-generated dataset discussed in the main text.
 }
 \label{fig:grad_mnisq} 
\end{figure}

\section{Justification for the low-rank approximation}
\label{sec:low_rank_justification}
\subsection{Theoretical guarantee on the classification risk}
In this section, we provide a theoretical justification for truncating the trained observable $O_{\boldsymbol{\alpha}, \mathcal{D}}$ defined in Eq.~\eqref{eq:implicit_model_ob} of the main text.
We compare upper bounds on the classification risk for the original classifier and its rank-$K$ truncation, and show that the deterioration is controlled by the magnitude of the first discarded eigenvalue, $\left|\lambda_{K+1}\right|$.

\subsubsection{Notation and setup}
Let $\rho(\boldsymbol{x})$ be a density operator, where the input $\boldsymbol{x}$ can represent classical data or serve as a label for quantum data.
Given a training sample $\{(\boldsymbol{x}_m,y_m)\}_{m=1}^{M}$, the $C$‑Support Vector Machines~(C-SVM) with regularization parameter $C>0$ learns the model parameter $\boldsymbol{\alpha}=(\alpha_1,\ldots,\alpha_M)^{\!\top}$, where each $\alpha_m$ corresponds to the product of a dual variable and the training label $y_m$.
The resulting trained observable is $O_{\boldsymbol{\alpha},\mathcal{D}}:=\sum_{m=1}^{M}\alpha_m\rho(\boldsymbol{x}_m)$.
Write the spectral decomposition
\begin{align}
    O_{\boldsymbol{\alpha},\mathcal D}
    =\sum_{i=1}^{r}\lambda_i
      \ket{\lambda_i}\bra{\lambda_i},
  \qquad
  |\lambda_1|\ge|\lambda_2|\ge\cdots\ge|\lambda_r|>0, 
\end{align}
where $\ket{\lambda_i}$ is the eigenvector corresponding to the eigenvalue $\lambda_i$, and $r$ is the rank of $O_{\boldsymbol{\alpha},\mathcal{D}}$.
The decision function of the original implicit model is 
\begin{align}
    f(\boldsymbol{x})=\operatorname{Tr}\left[O_{\boldsymbol{\alpha}, \mathcal{D}} \rho(\boldsymbol{x})\right].
\end{align}
For a prescribed rank parameter $K\le r$ we define the rank-$K$ approximation of the observable
\begin{align}
    O^{(K)}_{\boldsymbol{\alpha}, \mathcal{D}}:=\sum_{i=1}^{K}\lambda_i
          \ket{\lambda_i}\bra{\lambda_i},
\end{align}
and its associated decision function
\begin{align}
    f_K(\boldsymbol{x})
  :=\operatorname{Tr}\bigl[O^{(K)}_{\boldsymbol{\alpha}, \mathcal{D}}\rho(\boldsymbol{x})\bigr].
\end{align}
Using the bias $b\in\mathbb R$ chosen during training, we obtain binary classifiers for the original implicit model and its rank-$K$ approximation 
\begin{align}
    h(\boldsymbol{x})
    =\operatorname{sgn}\bigl(f(\boldsymbol{x})+b\bigr),
  \qquad
  h_{K}(\boldsymbol{x})
    =\operatorname{sgn}\bigl(f_K(\boldsymbol{x})+b\bigr).
\end{align}
For any classifier $g:\mathcal X\to\{-1,+1\}$, where $\mathcal X$ is the input data space, we define the classification risk as
\begin{align}
    R_{\mathrm{err}}(g)
  :=\mathbb{E}_{(\boldsymbol{x},y)\sim\mathcal P}
        \bigl[\mathds 1\!\bigl(y\neq g(\boldsymbol{x})\bigr)\bigr],
\end{align}
where $\mathcal P$ denotes the true (unknown) data distribution.
We use the hinge loss function $\phi_{\mathrm{hinge}}(z) = \max(0, 1-z)$.

\subsubsection{Main result}
To justify our low-rank approximation, we will prove the following key proposition, which bounds the classification risk of the truncated model.

\begin{proposition}
\label{prop:err_gap}
Let $(f,b)$ be the decision function trained by a C-SVM on a sample of size $M$, giving the classifier $h(\boldsymbol{x})
    =\operatorname{sgn}\bigl(f(\boldsymbol{x})+b\bigr)$.
Let $h_K$ be its rank-$K$ approximation.
Assume the kernel $k$ with $\sup_{\boldsymbol{x}\in \mathcal{X}}k(\boldsymbol{x},\boldsymbol{x})\le\Lambda^{2}$ and let the regularization parameter be $\mu_{M}>0$ (which is inversely related to the SVM hyperparameter $C$).
Then for any $K\le r$ and any $\delta\in(0,1)$, with probability at least $1-\delta$ over the training data $(\boldsymbol{x}_m,y_m)\sim\mathcal{P}^{\otimes M}$, the classification risk of the rank-$K$ truncated model $R_{\mathrm{err}}\left(h_K\right)$ is bounded by the standard upper bound for the classification risk of the original model $R_{\mathrm{err}}\left(h\right)$ plus an additional term $\left|\lambda_{K+1}\right|$ as follows:
\begin{align}
\label{eq:err_gap}
    R_{\mathrm{err}}\left(h_K\right) 
    &\leq
\frac{1}{M} \sum_{m=1}^M
\phi_{\mathrm {hinge }}\left(y_m\left(f(\boldsymbol{x}_m)+b\right)\right)
+\gamma_{M, \delta}
+\left|\lambda_{K+1}\right|,
\end{align}
where $\gamma_{M,\delta}:= 2(1+\sqrt{\log(2/\delta)}) \left(\frac{3\Lambda}{\sqrt{M\mu_{M}}} + \frac{2}{\sqrt{M}}\right)$ is a complexity term derived from standard uniform convergence bounds.
\end{proposition}

\subsubsection{Proof components}
The proof requires several intermediate lemmas and propositions. 
We will proceed by: 
\begin{enumerate}
    \item Bounding the difference between the original and truncated outputs in Lem.~\ref{lemma: function_output_bound}.
    \item Using this result to bound the change in hinge loss in Lem.~\ref{lemma:hinge}.
    \item Confirming that the truncated model remains in a well-behaved feasible set in Prop.~\ref{prop:feasible}. 
    \item Combining these with a standard uniform convergence bound to prove main result~\ref{prop:err_gap}.
\end{enumerate}

A key insight of our proof is that truncating the observable $O_{\boldsymbol{\alpha}, \mathcal{D}}$ in the feature space corresponds to an orthogonal projection of the decision function $f$ within the RKHS.

\begin{lemma}[Function output bound]
\label{lemma: function_output_bound}
The absolute difference between the decision function $f(\bm{x})$ and its truncated counterpart $f_K(\bm{x})$ is bounded as:

\begin{align}
    \left|f(\boldsymbol{x}) - f_K(\boldsymbol{x})\right| \leq \left|\lambda_{K+1}\right|
\end{align}
\end{lemma}

\begin{proof}
The difference in the function outputs is $\left|\operatorname{Tr}\left[(O_{\boldsymbol{\alpha}, \mathcal{D}} - O^{(K)}_{\boldsymbol{\alpha}, \mathcal{D}}) \rho(\boldsymbol{x})\right]\right|$. 
By applying Hölder's inequality, $|\operatorname{Tr}(AB)| \leq \|A\|_{\mathrm{op}}\|B\|_1$, and using the fact that $\|\rho(\boldsymbol{x})\|_1 = 1$ for any quantum state, this difference is bounded by $\|O_{\boldsymbol{\alpha}, \mathcal{D}} - O^{(K)}_{\boldsymbol{\alpha}, \mathcal{D}}\|_{\mathrm{op}}$.
Since $O_{\boldsymbol{\alpha}, \mathcal{D}} - O^{(K)}_{\boldsymbol{\alpha}, \mathcal{D}}=\sum_{i>K}\lambda_i\,|\lambda_i\rangle\langle\lambda_i|$, we have $\|O_{\boldsymbol{\alpha}, \mathcal{D}} - O^{(K)}_{\boldsymbol{\alpha}, \mathcal{D}}\|_{\mathrm{op}}=\left|\lambda_{K+1}\right|$, which proves the desired bound.
\end{proof}

\begin{lemma}[Hinge loss bound]
\label{lemma:hinge}
    For any $(\boldsymbol{x},y)\in\mathcal X\times\{-1,1\}$, the hinge loss $\phi_{\mathrm{hinge}}(z)$ satisfies:
    \begin{align}
        \phi_{\mathrm{hinge}}\bigl(y(f_K(\boldsymbol{x})+b)\bigr) \leq \phi_{\mathrm{hinge}}\bigl(y(f(\boldsymbol{x})+b)\bigr) + \left|\lambda_{K+1}\right|
    \end{align}
\end{lemma}

\begin{proof}
    The hinge loss is 1-Lipschitz continuous, meaning $|\phi_{\mathrm{hinge}}(a) - \phi_{\mathrm{hinge}}(b)| \leq |a - b|$. 
    Therefore,
    \begin{align}
        |\phi_{\mathrm{hinge}}\bigl(y(f_K(\boldsymbol{x})+b)\bigr) - \phi_{\mathrm{hinge}}\bigl(y(f(\boldsymbol{x})+b)\bigr)| 
        & \leq |y f_K(\boldsymbol{x}) - y f(\boldsymbol{x})|\\ 
        & = |f_K(\boldsymbol{x}) - f(\boldsymbol{x})|
    \end{align}
Finally, applying the bound from Lem.~\ref{lemma: function_output_bound} to the right-hand side and rearranging the terms yields the desired result.
\end{proof}

\begin{lemma}[Orthogonal projection shrinks the RKHS norm]
\label{lem:proj_nonexpansive}
Let $\mathcal H$ be any Hilbert space and let $P_{K}:\mathcal H\to\mathcal H$ be the orthogonal projection onto a $K$‑dimensional subspace $\mathcal H_{K}\subset\mathcal H$.  
Then for every $f\in\mathcal H$ one has  $\|P_{K}f\|_{\mathcal H}\le\|f\|_{\mathcal H}$.
\end{lemma}

\begin{proof}
Because $P_{K}$ is an orthogonal projection, it is self‑adjoint ($P_{K}=P_{K}^{*}$) and idempotent ($P_{K}^{2}=P_{K}$).  
Hence, by the Pythagorean theorem in Hilbert spaces,
\begin{align}
    \|f\|_{\mathcal H}^{2}
  =\|P_{K}f\|_{\mathcal H}^{2}
  +\|(\operatorname{Id}-P_{K})f\|_{\mathcal H}^{2}\ge
  \|P_{K}f\|_{\mathcal H}^{2},
\end{align}
which implies the claim.
\end{proof}

\begin{proposition}[Feasible set is projection‑invariant]
\label{prop:feasible}
Let $\mathcal H$ be the RKHS of a bounded kernel $k$ satisfying 
$\sup_{\boldsymbol{x}\in \mathcal{X}}k(\boldsymbol{x},\boldsymbol{x})\le\Lambda^{2}$ and fix a regularization parameter $\mu_{M}>0$.
Define
\begin{align}
    \mathcal G_{M}
  :=\Bigl\{(f,b)\in\mathcal H\times\mathbb R
      \,\big|\,\|f\|^{2}_{\mathcal H}\le1/\mu_{M},
                  \;|b|\le\Lambda/\sqrt{\mu_{M}}+1\Bigr\}.
\end{align}
For any $(f,b)\in\mathcal G_{M}$ and any orthogonal projection
$P_{K}:\mathcal H\to\mathcal H$ of rank~$K$, the pair
\((P_{K}f,\,b)\) also belongs to $\mathcal G_{M}$.
\end{proposition}

\begin{proof}
Lem.~\ref{lem:proj_nonexpansive} yields $\|P_{K}f\|_{\mathcal H}\le\|f\|_{\mathcal H}\le1/\sqrt{\mu_{M}}$, so the norm bound is preserved.
The bias $b$ is unchanged, hence its bound is preserved as well.
\end{proof}

The following result is a standard uniform‐deviation bound for C‑SVMs with bounded kernels~\cite{mohri2018foundations}; we state it in the exact form needed later.

\begin{theorem}[Uniform deviation of the hinge risk]
\label{thm:uc}
Assume the bounded‑kernel condition $\sup_{\boldsymbol{x}\in\mathcal{X}}k(\boldsymbol{x},\boldsymbol{x})\le\Lambda^{2}$ stated in Prop.~\ref{prop:feasible}, and let $\mathcal G_M$ be defined as in Prop.~\ref{prop:feasible}.
Define the composite class
$\phi_{\mathrm{hinge}}\circ\mathcal G_M
 :=\{     (\boldsymbol{x},y)\mapsto\phi_{\mathrm{hinge}}\!\bigl(y(f(\boldsymbol{x})+b)\bigr)
     \mid(f,b)\in\mathcal G_M\}$.
Then for any $\delta\in(0,1)$, with probability at least $1-\delta$ over an i.i.d.\ sample $\{(\boldsymbol{x}_m,y_m)\}_{m=1}^{M}\sim\mathcal P^{\otimes M}$,
\begin{align}
    \sup_{(f,b)\in\mathcal G_M}\Bigl|
    \mathbb E_{(X,Y)\sim\mathcal P}[\phi_{\mathrm{hinge}}\bigl(Y(f(X)+b)\bigr)]
    -\frac1M\sum_{m=1}^{M}\phi_{\mathrm{hinge}}\bigl(y_m(f(\boldsymbol{x}_m)+b)\bigr)
  \Bigr|
  \le 2(1+\sqrt{\log(2/\delta)})
       \Bigl(\frac{3\Lambda}{\sqrt{M\mu_{M}}}+\frac{2}{\sqrt{M}}\Bigr).
\label{eq:uc}
\end{align}
\end{theorem}

\subsubsection{Proof of the main proposition}
\label{proof:err_gap}
With all the necessary components in place, we now prove our main result.

\begin{proof}[Proof of Proposition~\ref{prop:err_gap}]
Because $h$ and $h_{K}$ are both sign predictors derived from $(f,b)$ and $(f_{K},b)$, respectively, we begin with the well‑known inequality $\mathds1(z\le0)\le\phi_{\mathrm{hinge}}(z)$ and write

\begin{align}
    R_{\mathrm{err}}(h)
   \;\le\;
   \mathbb{E}_{(X,Y)\sim\mathcal P}\bigl[\phi_{\mathrm{hinge}}\bigl(Y(f(X)+b)\bigr)\bigr],
\end{align}
with an analogous inequality for $h_{K}$.

Apply Th.~\ref{thm:uc} to both $(f,b)$ and $(f_{K},b)$.
Prop.~\ref{prop:feasible} guarantees that the truncated pair is admissible.  
Since Th.~\ref{thm:uc} is uniform over $\mathcal G_M$, it applies to both pairs without a union bound penalty, we obtain, simultaneously with probability $\ge1-\delta$,

\begin{align}
  R_{\mathrm{err}}(h)
  &\;\le\;
     \frac1M\sum_{m=1}^{M}
        \phi_{\mathrm{hinge}}\!\bigl(y_m(f(\boldsymbol{x}_m)+b)\bigr)
     +\gamma_{M,\delta},
  \label{eq:emp1}
  \\
  R_{\mathrm{err}}(h_{K})
  &\;\le\;
     \frac1M\sum_{m=1}^{M}
        \phi_{\mathrm{hinge}}\!\bigl(y_m(f_{K}(\boldsymbol{x}_m)+b)\bigr)
     +\gamma_{M,\delta},
  \label{eq:emp2}
\end{align}
where
$\gamma_{M,\delta}:=
  2(1+\sqrt{\log(2/\delta)})
  \bigl(\tfrac{3\Lambda}{\sqrt{M\mu_{M}}}
        +\tfrac{2}{\sqrt{M}}\bigr)$
matches the right‑hand side of~\eqref{eq:uc}.

Applying Lem.~\ref{lemma:hinge} to \eqref{eq:emp2}, we obtain
\begin{align}
    R_{\mathrm{err}}\left(h_K\right) 
    &\leq
\frac{1}{M} \sum_{m=1}^M
\phi_{\mathrm {hinge }}\left(y_m\left(f(\boldsymbol{x}_m)+b\right)\right)
+\gamma_{M, \delta}
+\left|\lambda_{K+1}\right|.
\end{align}
\end{proof}

\subsubsection{Discussion}
This proposition provides a theoretical foundation for our low-rank approximation strategy. 
Our goal is not to provide a tight estimate of the true classification risk, as uniform convergence bounds of this type are often loose in practice. 
Instead, the significance of our result lies in its comparative nature: we aim to rigorously quantify the effect of our approximation.

Our analysis proves that the certified performance guarantee for the truncated model $h_K$ deteriorates from the original model's guarantee for $h$ by an amount provably controlled by $\left|\lambda_{K+1}\right|$.
This linear error propagation is a direct consequence of the 1-Lipschitz continuity of the hinge loss function used in SVMs, which prevents the amplification of approximation errors. 
This makes our low-rank approximation not merely a heuristic, but a procedure with a provable bound on its error. 
The practical utility of this bound hinges on the condition that $|\lambda_{K+1}|$ is small. 
As we demonstrate in the following subsection (see Fig.~\ref{fig:add_analysis_b}), the eigenvalue spectra for the MNISQ dataset~\cite{Placidi2023-kq} and the VQE-generated dataset~\cite{Nakayama2023-zc} do indeed decay rapidly, providing strong empirical evidence that this condition is met.

This condition of a rapidly decaying spectrum is also consistent with the well-known manifold hypothesis~\cite{Tenenbaum2000,fefferman2013testingmanifoldhypothesis}.
This hypothesis suggests that many high-dimensional real-world datasets effectively lie on a much lower-dimensional manifold.
In the context of our model, this implies that the subspace spanned by the feature vectors, $\mathcal{S}= \operatorname{span}\left\{\left|\psi\left(\boldsymbol{x}_m\right)\right\rangle\right\}$, is itself low-dimensional.
As the trained observable $O_{\boldsymbol{\alpha}, \mathcal{D}}$ is a linear combination of $\rho\left(\boldsymbol{x}_m\right)$, its rank is upper-bounded by the dimension of $\mathcal{S}$.
The manifold hypothesis, therefore, provides a justification for why the trained observable in our framework is naturally expected to be highly amenable to low-rank approximation, making our approximation strategy effective.

\subsection{Empirical eigenvalue spectrum}
\label{sec:empirical_eigenvalue_spectrum}
\begin{figure}
 \centering
  \includegraphics[width=.7\linewidth]{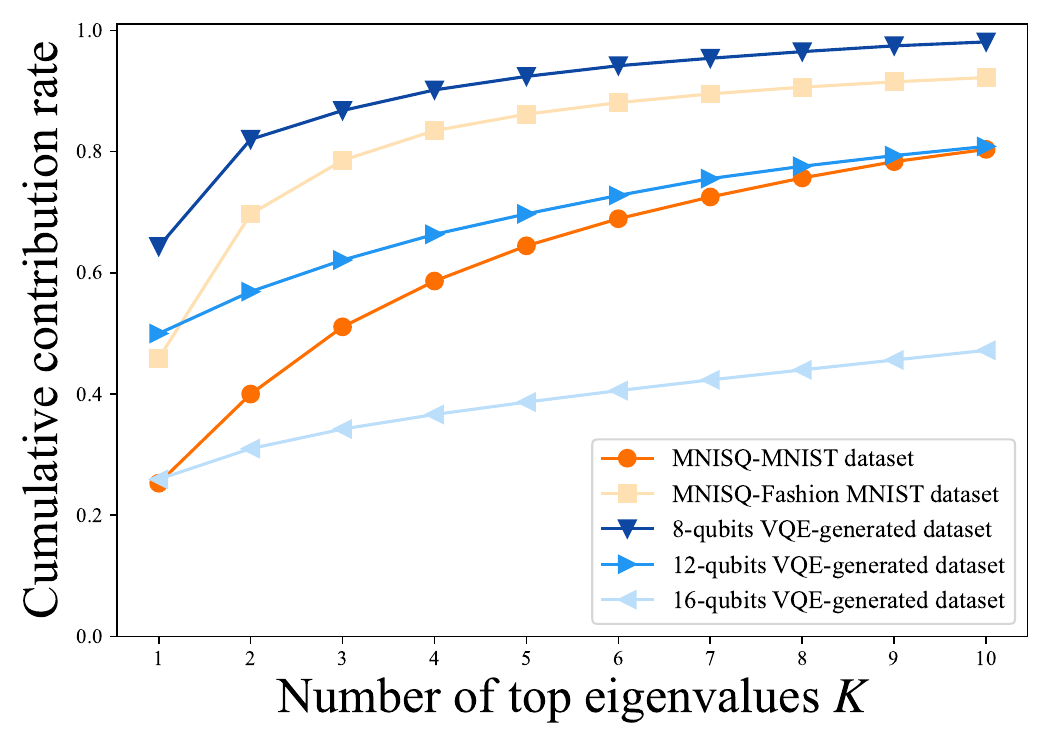}
 \caption{
The mean value of the cumulative contribution ratio, defined as $\frac{\sum_{i=0}^{K-1} \lambda_i^2}{\sum_{i=0}^{M-1} \lambda_i^2}$, for the observables of each implicit model.}
 \label{fig:add_analysis_b}
\end{figure}

In Fig.~\ref{fig:add_analysis_b}, we plot the cumulative contribution ratio of the eigenvalues, which empirically justifies our low-rank approximation strategy.

\section{Additional analysis}
\label{sec:additional_analysis}
\begin{figure}
 \centering
 \includegraphics[width=.7\linewidth]{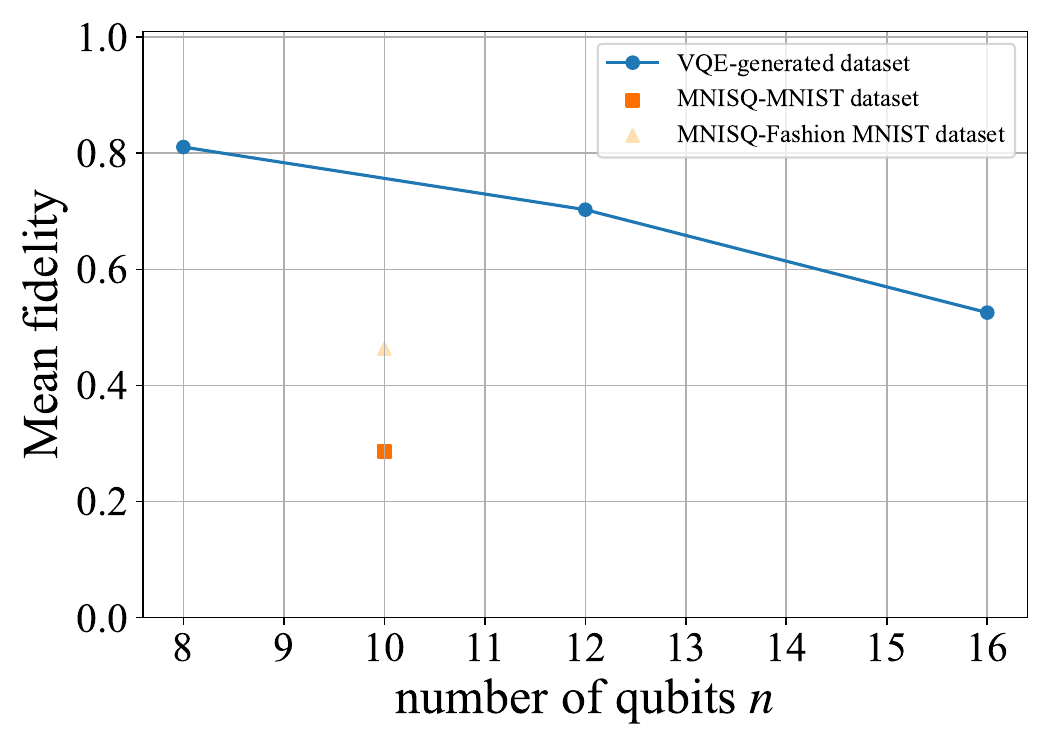}
 \caption{
Average fidelity between the first eigenvector of the observables of implicit models and the data for the corresponding distinguished classes.}
 \label{fig:add_analysis_a}
\end{figure}

\begin{figure}
 \centering
 \includegraphics[width=0.7\linewidth]{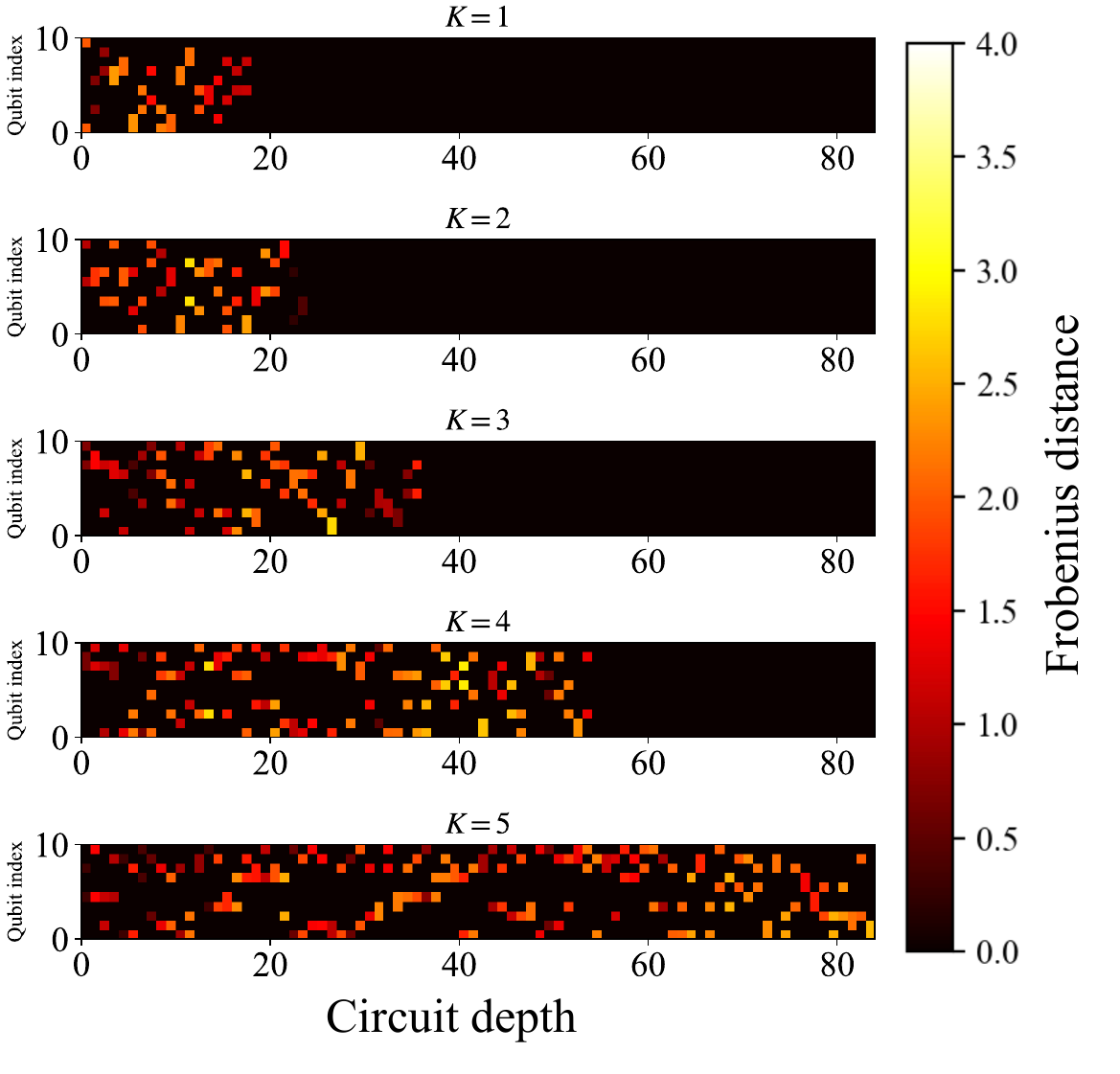}
 \caption{\textbf{A visualization of the quantum circuit that defines the EQS for MNISQ-MNIST dataset.}
 The vertical axis represents the index numbers of the qubits, while the horizontal axis corresponds to the depth of the quantum circuit.
 The color represents the Frobenius distance between the unitary gates and the identity matrix.}
 \label{fig:heat1}
\end{figure}

This section provides further analysis supporting the main text. 
In Fig.~\ref{fig:add_analysis_a}, we show the average fidelity between the first eigenvector of the implicit model's observable and data from the corresponding class.
Finally, Fig.~\ref{fig:heat1} shows an example of the quantum circuit structure of the EQS trained for label 0 of the MNISQ-MNIST dataset.

\section{Comparison with baseline classifiers}
\label{sec: baseline_comparison}
To place the performance of EQS in a broader context and further assess its practical advantage, we provide a comparative analysis against two simpler, quantum heuristic baseline classifiers, which use quantum feature states. 
The central question we address is whether the sophisticated construction of EQS is justified, or if a simpler heuristic can achieve comparable performance, particularly for datasets with high intra-class similarity.

The experiments use the same training and test splits of the MNISQ-MNIST datasets~\cite{Placidi2023-kq} and 12-qubit VQE-generated  datasets~\cite{Nakayama2023-zc} as used in the main text. 
We define two baselines.
The first baseline is a simple instance-based method inspired by the nearest neighbor rule~\cite{cover1967nearest,bishop2006pattern}, which we term a random sampling classifier. 
It assigns a new test state to the class with the highest average fidelity to $j$ representatives, randomly selected from the training data for each class. 
We report the mean and standard deviation over 10 independent trials for $j \in \{1, 5, 10\}$. 
Its prediction cost scales as $O(j)$.
The second baseline is the classic nearest centroid classifier~\cite{duda2006pattern}, implemented without explicitly constructing the centroid. 
Operationally, it classifies a new data point by calculating its average fidelity to all training states in each class and assigns it to the class with the highest score. 
Its prediction cost scales as $O(M)$.

The results of this comparison are summarized in Table~\ref{tab:baseline_comparison}. 
The table provides a nuanced picture. 
For the MNISQ-MNIST dataset, the EQS model outperforms both the random sampling and the nearest centroid classifiers. 
This suggests that for complex tasks, EQS inherits the sophisticated decision boundary from the trained SVM, going beyond simple class-average similarities. 
In contrast, for the highly-separable VQE-generated dataset, the simplest random sampling classifier achieves a higher accuracy than the trained SVM and its surrogate, EQS. 
This result suggests that the more complex SVM model overfitted on this simpler task, where the data's structure was highly separable.

Taken together, our results empirically demonstrate two key points: for complex tasks where a sophisticated kernel method excels, EQS provides a distinct performance advantage over simpler heuristics.
Furthermore, the VQE-generated dataset case highlights that the framework's power is unlocked only when surrogating an appropriately chosen, high-performing implicit model, making proper model selection an important prerequisite.

\begin{table}
\centering
\caption{
Comparison of EQS with baseline classifiers on test accuracy. 
The prediction cost $M$ denotes the size of the training set, while $j$ is the number of random samples.
The results for EQS are shown for $K=10$ on the MNISQ-MNIST and $K=6$ for the 12-qubit VQE-generated dataset.
}
\label{tab:baseline_comparison}
\begin{tabular}{l|c|c|c}
\hline\hline
\textbf{Method} & \textbf{Prediction Cost} & \textbf{MNISQ-MNIST Acc. (\%)} & \textbf{12-qubit VQE-generated Acc. (\%)} \\ \hline
\textbf{Baselines} & & & \\
\quad Random sampling ($j=1$) & $O(j)$ & $42.0 \pm 2.9$ & $95.4 \pm 4.4$ \\
\quad Random sampling ($j=5$) & $O(j)$ & $66.4 \pm 3.1$ & $99.1 \pm 0.3$ \\
\quad Random sampling ($j=10$) & $O(j)$ & $71.6 \pm 2.7$ & $99.2 \pm 0.1$ \\
\quad Nearest centroid & $O(M)$ & 79.0 & 99.4 \\
\hline
\textbf{Our work} & & & \\
\quad \textbf{EQS}  & \textbf{$O(1)$} & \textbf{93.8} & \textbf{81.2} \\
\hline
\textbf{Reference} & & & \\
\quad Original SVM & $O(M)$ & 95.2 & 81.9 \\
\hline\hline
\end{tabular}
\end{table}

\section{Discussion on applicability to other kernel types}
\label{sec: applicability_projected_kernels}
A natural direction for future research is to extend the EQS framework beyond the global fidelity quantum kernels used in this work to other types of quantum kernels. 
This would allow EQS to tackle a broader class of problems by leveraging models with different inductive biases.
Constructing EQS involves two primary steps: (1) efficiently diagonalizing the trained observable, and (2) constructing a quantum circuit that produces the obtained eigenvectors. 
The feasibility of extending EQS depends on both steps, but their challenges are distinct. 
The feasibility of the second step is not dependent on the kernel type itself (once the eigenvectors are given). 
In contrast, the feasibility of the first step depends heavily on the chosen kernel's structure.
Therefore, efficiently diagonalizing the trained observable is the key challenge, and the primary bottleneck, for extending the EQS framework.

As a notable relevant example, we analyze the extension to projected quantum kernels~\cite{gan2023unified}. 
When their inherent inductive bias aligns well with the problem's underlying structure, they can offer a mechanism to sidestep the curse of dimensionality, potentially enabling learning on datasets where the global fidelity quantum kernel might fail due to exponential concentration effects~\cite{thanasilp2024exponential}.
To clarify this discussion, we analyze the feasibility of this extension for two common types of linear projected quantum kernels (LPQKs)~\cite{gan2023unified}: the $\mathbf{s}$-LPQK and $S$-LPQK. 

First, we consider $\mathbf{s}$-LPQK.
We define the reduced density matrix as $\rho_{\mathbf{s}}(\boldsymbol{x})=\operatorname{tr}_{\bar{\mathbf{s}}}(\rho(\boldsymbol{x}))$, where $\mathbf{s}$ denotes the set of $S$ qubit indices specifying the subsystem (i.e., $S=|\mathbf{s}|$).
The $\mathbf{s}$-LPQK is defined on this single subsystem:
\begin{align}
    k_{\mathbf{s}}\left(\boldsymbol{x}, \boldsymbol{x}^{\prime}\right)=\operatorname{Tr}_{\mathbf{s}}\left(\rho_{\mathbf{s}}(\boldsymbol{x}) \rho_{\mathbf{s}}(\boldsymbol{x}')\right).
\end{align}
The associated implicit model is
\begin{align}
\label{eq:s-lpqk_model}
    f_\mathbf{s}(\boldsymbol{x})=\sum_{m=1}^M \alpha_m k_\mathbf{s}\left(\bm{x}_m, \bm{x}\right)=\operatorname{Tr}\left[O_\mathbf{s} \rho_\mathbf{s}(\bm{x})\right],
\end{align}
where $O_{\mathbf{s}}=\sum_m^M \alpha_m \rho_{\mathbf{s}}\left(\boldsymbol{x}_m\right)$.

The core task is diagonalization of the trained observable $O_{\mathbf{s}}$.
This task is fundamentally different from the global fidelity quantum kernel case.
For the global fidelity quantum kernel, the observable $O_{\mathrm{global }}=\sum_{m=1}^M \alpha_m \rho(\boldsymbol{x}_m)$ is a sum of pure states, guaranteeing its eigenvectors lie within an at most $M$-dimensional subspace spanned by the training feature states, $\operatorname{span}\left\{\left|\psi\left(\boldsymbol{x}_m\right)\right\rangle\right\}_{m=1}^M$. 
This makes it computationally efficient to diagonalize.
For the $\mathbf{s}$-LPQK, the eigenvectors can span the entire $2^S$-dimensional Hilbert space.
However, this is expected to be computationally feasible, provided $S$ is small, which is the exact regime where $\mathbf{s}$-LPQKs are employed.

Once $O_{\mathbf{s}}$ is obtained and diagonalized as a classical $2^S \times 2^S$ matrix, we also obtain the classical representation of the unitary $U_{\mathbf{s}}$ that maps the computational basis to the eigenvectors ($U_{\mathbf{s}}|k\rangle = |\lambda_k\rangle$).
From this explicit matrix, the corresponding quantum circuit $\mathcal{C}$ can be analytically constructed using well-established classical compilation algorithms~\cite{1629135, mottonen2004quantum}.
It is worth noting that AQCE is also applicable in principle. 
The feasibility of AQCE hinges on the ability to efficiently compute the fidelity tensor~\eqref{eq:fidelity_tensor_original}, which fundamentally relies on estimating inner products involving the target eigenvectors $|\lambda_k\rangle$. 
As long as these inner products can be efficiently estimated (which is expected given $S$ is small), AQCE remains a viable pathway for the circuit construction.

Next, we consider the $S$-LPQK, which is defined as a sum of $\mathbf{s}$-LPQK over all possible subsystems of size $S=|\mathbf{s}|$. 
Let $\mathbb{S}_S=\left\{\mathbf{s}_1, \mathbf{s}_2, \ldots, \mathbf{s}_W| | \mathbf{s}_i \mid=S\right\}$ be the set of all $W=\binom{n}{S}$ possible $S$-qubit subsystems. 
The $S$-LPQK is defined as:
\begin{align}
k_S\left(\boldsymbol{x}, \boldsymbol{x}^{\prime}\right)=\frac{1}{\sqrt{W}} \sum_{\mathbf{s} \in \mathbb{S}_S} k_{\mathbf{s}}\left(\boldsymbol{x}, \boldsymbol{x}' \right).
\end{align}
Its implicit model is
\begin{align}
    f_S(\boldsymbol{x}) &=\sum_{m=1}^M \alpha_m k_S\left(\bm{x}_m, \bm{x}\right) = \operatorname{Tr}\left[O_S \rho(\boldsymbol{x})\right],
\end{align}
where $O_S:=\frac{1}{\sqrt{W}} \sum_{\mathbf{s} \in \mathbb{S}_S}\left\{\left(\sum_m^M \alpha_m \rho_{\mathbf{s}}\left(\boldsymbol{x}_m\right)\right) \otimes I_{\bar{\mathbf{s}}}\right\}$.

Diagonalizing the trained observable $O_S$ presents a different challenge than the $\mathbf{s}$-LPQK case.
The observable $O_S$ is an $n$-qubit operator, and its diagonalization is generally computationally prohibitive as it acts across the full $2^n$-dimensional space.
However, in special cases where the observable $O_S$ is efficiently diagonalizable, we can also extend the EQS framework.
A simple example of such a scenario occurs if the set $\mathbb{S}_S$ is restricted to a collection of non-overlapping subsystems.
In this case, the operators that constitute $O_S$ are automatically mutually commutative, and the unitary circuit $\mathcal{C}$ that simultaneously diagonalizes them decomposes into a tensor product of local circuits (i.e., $\mathcal{C} = \mathcal{C}_{\mathbf{s}_1} \otimes \mathcal{C}_{\mathbf{s}_2} \otimes \dots \otimes I_{\mathrm{rest}}$ for a non-overlapping set $\mathbb{S}_S = \{\mathbf{s}_1, \mathbf{s}_2, \dots\}$).
Each local circuit $\mathcal{C}_\mathbf{s}$ diagonalizes its corresponding $O_\mathbf{s}$ and can be constructed classically with a cost equivalent to the $\mathbf{s}$-LPQK case.
The $S=1$ case is the simplest example of this.

In summary, extending EQS to projected kernels is a feasible strategy for the $\mathbf{s}$-LPQK.
The extension to more general $S$-LPQKs remains future work; overcoming this computational hurdle is the research effort required to unlock the full potential of EQS.

\end{document}